\documentclass[12pt]{article}
\usepackage{graphicx}
\usepackage{dcolumn}
\usepackage{bm}
\usepackage{graphics,exscale}
\usepackage{amsmath}
\usepackage{amssymb}
\usepackage{amscd}
\usepackage{afterpage}
\usepackage{float,times}
\usepackage{subfigure}
\usepackage{rotating}
\usepackage{multirow}
\usepackage{fancyheadings}
\usepackage{epsfig}
\usepackage{theorem}
\usepackage{moreverb}
\usepackage{euscript}
\usepackage{psfrag}

\newcommand{\ads}{$AdS_7$\ }
\newcommand{\ad}{$AdS_9$\ }
\newcommand{\g}{\mathcal{G}}

\begin{document}
\pagestyle{empty}
{\hbox to\hsize{\hfill October 2009 }}

\vspace*{10mm}
\begin{center}

{\Large\bf Warping, Extra Dimensions and a Slice of $\mathbf{AdS_d}$}\\
\vspace{1.0cm}

{\large Kristian L. McDonald\footnote{Email: klmcd@triumf.ca}}\\
\vspace{1.0cm}
{\it {Theory Group, TRIUMF, 4004 Wesbrook Mall, Vancouver, BC V6T2A3, Canada.}}\\
\vspace{1.4cm}

\end{center}
\begin{abstract}
Inspired by the Randall-Sundrum (RS) framework we consider a number of phenomenologically relevant model building questions on a slice of compactified $AdS_d$ for $d >5$. Such spaces are interesting as they enable one to realize the weak scale via warping. We perform the Kaluza-Klein (KK) reduction for gravitons and bulk vectors in these spaces and for the case of $AdS_6$ consider the KK spectrum of gauge-scalars. We further obtain the KK towers for bulk fermions on a slice of $AdS_7$ and $AdS_9$ and show that the RS approach to flavor generalizes to these spaces with the localization of chiral zero mode fermions controlled by their bulk Dirac mass parameters. However for the phenomenologically interesting case where the transverse radius is $R^{-1}\sim$~TeV we show that bulk Standard Model fields are not viable due to a resulting volume suppression of the gauge coupling constants. A similar suppression occurs for the case of UV localization. Thus it seems that the Standard Model fields should be confined to the infrared brane in such spaces. Sterile fields and extended gauge sectors may propagate in the bulk with the gauge-coupling volume suppression experienced by the latter motivating a weak coupling to Standard Model fields. We also discuss some issues regarding the effective 4D theory description in these spaces.
\end{abstract}

\vfill

\eject
\pagestyle{empty}
\setcounter{page}{1}
\setcounter{footnote}{0}
\pagestyle{plain}
\section{Introduction}
Though a remarkably successful theory, the Standard Model (SM) of particle physics is almost certainly incomplete. There are two main reasons, one theoretical the other experimental, that lead us to suspect that new physics will appear at the TeV scale. The direct sensitivity of the Higgs mass to ultraviolet (UV) effects (the hierarchy problem) makes it difficult to take the SM seriously as a successful theory beyond the TeV scale. A likely scenario is that some mechanism is responsible for stabilizing the weak scale and the expectation is that this mechanism will manifest itself in the form of new particles with $\sim$~TeV masses. On the experimental side there is now a growing body of evidence suggesting that the matter density of the universe is dominated by an unknown particle or particles, referred to as dark matter (DM). Curiously the requisite behaviour of the DM can be obtained by a $\sim10^2$~GeV particle which interacts with weak scale strength with the SM fields. 

One promising possibility is that Nature is supersymmetric, in which case there should exist $\sim$~TeV scale particles whose UV sensitive contribution to the Higgs mass via loop effects approximately cancel the UV sensitive contributions of SM particles. Supersymmetric extensions of the SM can also motivate the unknown DM density as the imposition of an extended symmetry ($R$-parity) on supersymmetric models renders the lightest new particle absolutely stable. Furthermore the coupling constant relations dictated by supersymmetry mandate weak scale interaction strengths for some of the supersymmetric particles.

An alternative solution to the hierarchy problem can occur if Nature possesses extra spatial dimensions. In particular if nature admits a non-factorizable geometry the weak scale may be realized as a red-shifted, or warped, incarnation of Planck scale sized input parameters~\cite{Randall:1999ee}. In this case the break down of the SM at the TeV scale would be manifest by the existence of TeV scale Kaluza-Klein (KK) excitations of the graviton, and of the SM fields, if the latter propagate in the bulk. Interestingly if SM fermions propagate in the bulk theories of flavor can also be constructed by employing the wave function overlap of the SM fermions in the extra space~\cite{Grossman:1999ra,Gherghetta:2000qt}. The existence of extra spatial dimensions can also motivate a DM candidate if a subgroup of an isometry of the extra space is conserved in the low energy theory. This is precisely what happens in models with universal extra dimensions (UED models)~\cite{Appelquist:2000nn} where all the SM fields propagate in the bulk of an extended spacetime and a remnant discrete symmetry, known as KK parity\footnote{KK parity may also be imposed on RS models by gluing together multiple warped throats~\cite{Agashe:2007jb}.}, renders the lightest KK particle a good DM candidate~\cite{Servant:2002aq}. 

If the RS scenario is realized in nature it is possible that additional spatial dimensions exist beyond the warped extra dimension. As discussed in~\cite{Bao:2005ni}, from a string theoretic perspective one may obtain the $AdS_5$ RS model from a stack of parallel $D3$ branes in type-IIB string theory~\cite{Aharony:1999ti}, though additional compact dimensions will be present. Interestingly one may realize $AdS_7$ with additional compact dimensions from a stack of parallel $M5$ branes in $M$ theory~\cite{Aharony:1999ti}. It has also been noted that $AdS_6$ (with additional compact dimensions) is the near-horizon limit of the Type I' D4-D8 brane system~\cite{Brandhuber:1999np}. It is important to ask how these extra dimensions, if present, may modify our understanding of the RS model and what new features may emerge.

In a recent work we have considered the generalization of the RS model to the higher dimensional space $AdS_5\times T^2$~\cite{McDonald:2009md}. In that work we were primarily motivated by the observation that UED models and RS models are, in some sense, complementary. RS models motivate the weak/Planck hierarchy, the existence of TeV scaled particles (in the form of KK excitations) and can shed light on the flavor puzzle. However the warped geometry breaks translational invariance along the extra dimension in a maximal fashion so that the KK particles are not stable and do not admit a good DM candidate. UED models on the other hand motivate a stable DM candidate but do not shed any light on the weak/Planck hierarchy, nor do they provide any insight into the flavor structure of the SM\footnote{Flavor structures may be viable in UED variants like split-UED~\cite{Park:2009cs}.}. The presence of the UED DM candidate at the weak scale is also not motivated within UED models as one obtains the TeV scale DM particle simply by assuming that the weak scale is similar to the KK scale; two scales which are otherwise independent. In~\cite{McDonald:2009md} we showed that the extended space $AdS_5\times T^2$ permits the complementary features of the RS and UED frameworks to be unified, with the warped direction motivating the weak/Planck hierarchy and admitting a description of flavor whilst KK parity emerges as a remnant symmetry of the extra toroidal dimensions. Interestingly the $AdS_5$ warping also motivates the connection between the weak scale and the UED KK scale with the warping inducing an effective KK scale on the torus of order $\sim$~TeV, even if the toroidal scale is $R^{-1}\sim M_{Pl}$. This motivates the connection between the weak scale and the DM scale usually assumed in UED models.

In the present work we extend the program undertaken in~\cite{McDonald:2009md} and consider the promotion of the RS model to a higher dimensional slice of compactified $AdS_d$ for $d>5$. Our motivations are ultimately phenomenological and we seek to determine the extent to which the complementary features of UED and RS models can be combined in these higher dimensional warped spaces. However there is also a theoretical aspect to our work as we generalize many familiar RS expressions to these higher dimensional warped spaces. We find that, as one would expect, the RS realization of the weak scale via spacetime warping carries over to a slice of $AdS_d$ for $d>5$ when the Higgs boson is a $(d-1)$ dimensional field localized on the IR brane. We further find that the RS approach to flavour also carries over to \ads and $AdS_9$, with the coupling between two chiral zero mode fermions and a brane localized scalar being exponentially sensitive to the fermion bulk mass parameters such that hierarchical Yukawas are expected in the 4D theory. However for the phenomenologically interesting case of $R^{-1}\sim1$~TeV the effective 4D gauge coupling between a chiral zero mode fermion and the zero mode of a bulk gauge field experiences volume suppression and, if the IR brane scale is $\sim$~TeV, the effective gauge coupling in the 4D theory is significantly suppressed. The severity of this suppression increases with $d$, though already for $d=7$ it is of order $\sim 10^{-15/2}$. Consequently bulk SM fermions and gauge fields are not viable for both \ads and $AdS_9$. The appealing RS approach to flavour is therefore viable only on a slice of $AdS_5$ or, as shown in~\cite{McDonald:2009md} for the case of $AdS_5\times T^2$, for certain spaces of the form $AdS_5\times \mathcal{M}^{d-5}$. 

If the SM matter fields propagate in the transverse dimensions of $AdS_d$ one therefore expects them to be localized at either the UV or infrared (IR) brane. We shall show that in the former case a similar suppression of the effective 4D couplings is found for $R^{-1}\sim$~TeV so that only IR localization is viable. The main model building feature of the $AdS_d$ spaces seems to be their ability to combine the warped explanation for the weak/Planck hierarchy with the KK parity found in UED models, so these spaces admit only a partial unification of the appealing complimentary features of UED and RS models. The main experimental signature for the $AdS_d$ spaces in this instance is the observation of warped KK gravitons in addition to UED KK modes. Such a signature also occurs when the $(d-1)$ dimensional UED model is realized by embedding the SM fields on the IR brane of $AdS_5\times T^{d-5}$, as discussed in~\cite{McDonald:2009md}. However, as we shall show, the graviton KK towers on $AdS_d$ and $AdS_5\times T^{d-5}$ differ so that if a $(d-1)$ dimensional UED scenario is discovered one would be able to experimentally determine if the UED model in is embedded in either of these distinct warped spaces by carefully studying the graviton KK spectrum.

Fermions which are sterile with respect to the SM gauge group may propagate in the bulk, with such a scenario considered already for \ads in the context of a brane localized UED model in~\cite{Appelquist:2002ft}. Extended gauge sectors may also propagate in the bulk and the resulting volume suppression of the gauge coupling can motivate a very weak coupling for such sectors with SM fields. For example, if the SM is localized on the IR brane of $AdS_6$, to realize an embedding of the minimal UED model on the IR brane and simultaneously motivate the weak/Planck hierarchy, the gauge group extension $\mathcal{G}_{SM}\times \mathcal{G}_X$ with $\mathcal{G}_X$ in the bulk permits the $\mathcal{G}_X$-symmetry breaking to occur on the IR brane at the weak scale and yet remain experimentally viable. Such a scenario may offer an interesting way to employ, for example, a weakly coupled symmetry which plays a custodial role and is broken at the weak scale. 

Before proceeding we note that works based on higher dimensional warped spaces exist already in the literature; see for example~\cite{Gherghetta:2000qi,Chacko:1999eb,Dubovsky:2000av,Gherghetta:2002xf}. It is known, for example, that in $AdS_7$ the cancellation of boundary anomalies~\cite{Gherghetta:2002xf} necessarily constrains the boundary symmetries and field content. The combination of warped and universal extra dimensions has been previously considered on a slice of $AdS_7$~\cite{Appelquist:2002ft} and the graviton KK tower for $AdS_7$ was also studied in~\cite{Bao:2005ni}. Some matters regarding moduli stabilization via bulk scalar fields in higher dimensional warped spaces were considered in~\cite{Flachi:2003bb} and the Casimir force was studied in~\cite{Saharian:2003qs} and ~\cite{Frank:2008dt}, where, in the latter, it was noted that the contribution from the transverse extra dimensions resembles that of UED models. A study of DM candidates that result from approximate isometries of warped throats in compactified string models has also appeared~\cite{Frey:2009qb}.

The layout of the present work is as follows. In Section~\ref{sec:grav} we consider the Einstein equations and graviton KK tower for $AdS_d$ and in Section~\ref{sec:vec_ads_d} we obtain the KK tower for bulk vectors in said spaces. Relative to $AdS_5$ the spaces $AdS_d$ for $d>5$ admit additional modes in the form of metric and gauge boson polarizations in the transverse space. As an example of these modes we detail the KK spectrum for gauge-scalars in Section~\ref{sec:vec_ads_d} for the $d=6$ case of $AdS_6$; six being the lowest dimensionality which admits such modes. We derive the KK spectra for bulk fermions on a slice of $AdS_7$ and $AdS_9$ in Section~\ref{sec:ads7_kk_fermi} and show that in each case a single localizable chiral zero mode appears in the spectrum. In Section~\ref{sec:couplings} we combine a number of these ingredients and consider the realization of the weak scale via warping with an IR brane localized Higgs boson, the mechanism of 4D flavor via fermion wavefunction overlap with an IR brane Higgs for $AdS_7$ and $AdS_9$ and the coupling of bulk vectors to bulk fermions in these spaces. Finally we comment on the range of validity of the effective 4D theory description and the case of UV localization in Section~\ref{sec:eff_4d_theory} before concluding in Section~\ref{sec:conc}. In four Appendices we provide additional information which complements the analysis, including our conventions for bulk fermions in 7D and 9D.
\section{Gravity on $\mathbf{AdS_d}$\label{sec:grav}}
We consider the metric defined by the $d$-dimensional spacetime interval
\begin{eqnarray}
ds^2&=&e^{-2\sigma(y)}\left[\eta_{\mu\nu}dx^\mu dx^\nu -\delta_{ab}dx^{a}dx^{b}\right]-dy^{2}\equiv G_{MN}dx^Mdx^N\label{ads6_metric},
\end{eqnarray}
where $M,N=0,1,2,3,5,6,..,d$ label the full $d$-dimensional space, $\mu,\nu=0,1,2,3$ label the 4D subspace and the extra dimensions are labeled by $x^{a}$, with $a,b=5,6,..,(d-1)$, and $x^d=y$ (the latter being the warped direction). The extra dimensions are compact with $x^{a}\in[-\pi R,\pi R]$, $y\in[-\pi r_c,\pi r_c]$, and the points $x^{a}=\pm\pi R $ ($y=\pm \pi r_c$) identified. For simplicity we take equal radii in the $x^{a}$ directions and we shall, at times, refer to these as the `transverse' extra dimensions. As in the RS model the warped direction is orbifolded as $S^1/Z_2$ with the $Z_2$ action defined by the identification $Z_2: y\rightarrow -y$. The transverse directions must also be orbifolded to ensure the absence of massless gravi-vectors. For much of what follows we need not specify this orbifolding, though for completeness we note that for odd $(d-5)$ we shall use
\begin{eqnarray}
(T^2/Z_2\times ....\times T^2/Z_2)\times S^1/Z_2,\label{ads_d_orbifolding}
\end{eqnarray}
where there are $(d-6)/2$ factors of $T^2/Z_2$ in the brackets. For even $(d-5)$ the last factor of $S^1/Z_2$ in (\ref{ads_d_orbifolding}) is not present and there are $(d-5)/2$ factors of $T^2/Z_2$. We provide additional details regarding this orbifolding as appropriate in the text. 

We take as sources a cosmological constant $\Lambda$ and two codimension one branes with tensions $V_{0,L}$; the resulting Einstein equations being
\begin{eqnarray}
& &\sqrt{G}\left[R_{MN}-\frac{1}{2}G_{MN}R^{(d)}\right]=\nonumber\\
&&-\frac{1}{4M_*^{d-2}}\left[\sqrt{G}G_{MN}\Lambda+\delta^{\bar{M}}_M\delta^{\bar{N}}_N\sqrt{\bar{G}}\bar{G}_{\bar{M}\bar{N}}\left\{V_0\delta(y)+V_L\delta(y-\pi r_c)\right\}\right].
\end{eqnarray}
Here $M_*$ ($R^{(d)}$) is the $d$-dimensional Planck scale (Ricci scalar), $\bar{G}_{\bar{M}\bar{N}}$ denotes the induced five dimensional metric at the brane locations with brane Lorentz indices $\bar{M},\bar{N}=0,1,2,3,5,6,..,(d-1)$, and $G=|\mathrm{det}(G_{MN})|$ (similarly for $\bar{G}$). The Einstein equations give
\begin{eqnarray}
(d-1)(d-2)\frac{\sigma'^2}{2}&=&-\frac{1}{4M_*^{d-2}}\Lambda,\label{E55}\\
(d-2)\sigma''&=&\frac{1}{4M_*^{d-2}}\left\{V_0\delta(y)+V_L\delta(y-\pi r_c)\right\},\label{Emn}
\end{eqnarray}
with solution
\begin{eqnarray}
\sigma&=&\sqrt{\frac{-\Lambda}{2(d-1)(d-2)M_*^{d-2}}}|y|\equiv k|y|,\label{sigma_sol}
\end{eqnarray}
so the warp factor may be written as $e^{-\sigma}=e^{-k|y|}$. Calculating the second derivative of $\sigma$ and comparing with (\ref{Emn}) requires the tunings
\begin{eqnarray}
V_0=-V_L=8(d-2)kM_*^{d-2},\label{tension_tune}
\end{eqnarray}
and the effective 4D Planck scale is given by
\begin{eqnarray}
M_{Pl}^2=\frac{2}{(d-3)}\frac{M_*^{d-2}}{k}(2\pi R)^{d-5}\left\{1-e^{-(d-3)k\pi r_c}\right\}.\label{4d_planck}
\end{eqnarray}
We note that the solution (\ref{sigma_sol}) requires $\Lambda<0$ and also define a new (conformal) variable by $kz=e^{ky}$ to write (\ref{ads6_metric}) as
\begin{eqnarray}
ds^2&=&\frac{1}{(kz)^2}\left[\eta_{\mu\nu}dx^\mu dx^\nu -\delta_{ab}dx^{a}dx^{b}-dz^2\right].\label{ads_conformal_metric}
\end{eqnarray}
We refer to this metric, sourced by a negative cosmological constant, as $AdS_d$. Strictly speaking the compactification of the transverse dimensions breaks some of the isometries usually present in $AdS_d$ (see~\cite{Gibbons:1998th}) and our space is the compactification of $AdS_d$ by the action of the discrete translation isometries $x^a\sim x^a+ 2\pi R$. For brevity we refer to this simply as $AdS_d$ with the implied compactification understood. 

We display the approximate value of the 7D gravity scale $M_*$ in Table~\ref{table:ads_d grav_scale} for\footnote{We note that the more extreme case of $d=32$ gives $M_*\sim $~TeV. In such a scenario the KK copies of SM fields would act as the extra sectors discussed already in connection with the hierarchy problem in, e.g.,~\cite{Dvali:2009ne}. We do not consider such large values of $d$ in this work and instead restrict our attention to $d<10$.} $d\in [5,9]$. Throughout this work we take $R^{-1}\sim$~TeV as this is the interesting region to be explored by the LHC and is also the compactification scale for UED models which permits the lightest KK particle to be a suitable DM candidate. We also take the IR brane scale as $\sim$~TeV so the hierarchy  between the fundamental gravity scale $M_*$ and the weak scale results from warping. For the phenomenologically interesting case of $R^{-1}\sim$~TeV with $k\sim M_*$ equation (\ref{4d_planck}) gives $(M_*/\mathrm{TeV})\sim 10^{30/(d-3)}$. As can be seen in the table, $M_*$ decreases with increasing $d$ for $R^{-1}\sim$~TeV due to the relatively large transverse volume. The IR brane scale is $e^{-k\pi r_c}M_*$ so that the value of $kr_c$ required to realize the weak scale on the IR brane also decreases with $d$. We also show this in the Table. One observes that, as opposed to the RS value of $kr_c \sim\mathcal{O}(10)$, no hierarchy is required for larger values of $d$ with $k r_c \sim\mathcal{O}(1)$ readily obtained.
\begin{table}[ht]
\centering
\begin{tabular}{|c|l|l|l|l|l|}
\hline
$d$&5&6&7&8&9\\ \hline
$\sim M_*/\mathrm{TeV}$& $10^{15}$&$10^{10}$& $10^{15/2}$& $10^{6}$& $10^{5}$ \\\hline
$\sim k r_c$&$11$&$7.3$&$5.5$&$4.4$&$3.7$\\

\hline
\end{tabular}
\caption{Approximate value of the 7D gravity scale $M_*$ and the warping parameters $kr_c$ for $AdS_d$ in the phenomenologically interesting case of $R^{-1}\sim$~TeV with $k\sim M_*$. Note that $M_*$ decreases with increasing $d$ and that $\mathcal{O}(1)$ values of $kr_c$ are allowed for larger $d$.}
\label{table:ads_d grav_scale}
\end{table}
\subsection{Graviton KK spectrum}
The masses and wave functions of the KK gravitons are found by making the metric replacement $G_{\mu\nu}=e^{-2\sigma}\eta_{\mu\nu}\rightarrow e^{-2\sigma}(\eta_{\mu\nu}+\kappa h_{\mu\nu})$, where $\kappa=2 M_*^{-(d-2)/2}$. The KK expansion for $h_{\mu\nu}$ is:
\begin{eqnarray}
h_{\mu\nu}(x^\sigma,x^a,z)=\sum_{\vec{n}}h^{(\vec{n})}_{\mu\nu}(x^\sigma)g^{(n_a)}_+(x^a)f_h^{(\vec{n})}(z),\label{h_kk}
\end{eqnarray}
where\footnote{We emphasize that $n_a$ ($n$) labels the quantized momenta in the compact $x^a$ ($z$) directions. We shall on occasion also denote $f^{(\vec{n})}_h$ as $f^{(n_a,n)}_h$.} $\vec{n}=(n_a,n)=(n_5,n_6,..,n_{d-1},n)$ and $g^{(n_a)}_+(x^a)$ are even parity wave functions on the transverse space. Working in the gauge $\partial^\mu h_{\mu\nu}=h^\mu_\mu=0$ the expansion (\ref{h_kk}) leads to
\begin{eqnarray}
\left[z^2\partial_z^2-(d-2)z\partial_z+m_{h,\vec{n}}^2z^2-m_{n_a}^2z^2\right]f^{(\vec{n})}_h=0,\label{grav_eom_f}
\end{eqnarray}
where we write the KK masses as $m_{h,\vec{n}}$ and use $\sum_a\partial_a^2g^{(n_a)}_+= -m_{n_a}^2g^{(n_a)}_+$. The profiles obey the orthogonality conditions
\begin{eqnarray}
\int  \frac{dz}{(kz)^{(d-2)}}f_h^{(n_a,n)}f_h^{(n_a,m)}&=&\delta^{nm},\nonumber\\
\int [\Pi_adx^a]g^{(n_a)}_+g^{(n_b)}_+&=&\delta^{n_an_b},
\end{eqnarray}
and the solutions to (\ref{grav_eom_f}) are
\begin{eqnarray}
f_h^{(\vec{n})}(z)=\frac{(kz)^{\nu_h}}{N_{\vec{n}}}\left\{J_{\nu_h}\left[\sqrt{m_{h,\vec{n}}^2-m_{n_a}^2}z\right]+\beta_{\vec{n}}Y_{\nu_h}\left[\sqrt{m_{h,\vec{n}}^2-m_{n_a}^2}z\right]\right\},\label{grav_profile}
\end{eqnarray}
where $N_{\vec{n}}$ is a normalization factor,  $\beta_h^{(\vec{n})}$ is a constant and the order of the Bessel functions is
\begin{eqnarray}
\nu_h=\frac{1}{2}(d-1).
\end{eqnarray}
Equation (\ref{grav_profile}) is the generalization of the RS ($AdS_5$) result and as such reduces to known expressions in the literature; the $d=5,6,7$ cases reproduce the  $AdS_{5,6,7}$ results found in references~\cite{Davoudiasl:1999jd},~\cite{Gherghetta:2000qi} and~\cite{Bao:2005ni} respectively\footnote{For string theoretic realizations there may be additional winding modes present in the spectrum. These can be phenomenologically important (see~\cite{Bao:2005ni}), though we do not consider them here.}. Note that for $d>5$ the profiles along the warped direction differ from the $d=5$ RS result with both the order of the Bessel functions and the power of the the prefactor $(kz)$ increasing.  That the warped wave functions for $AdS_d$ do not match those of $AdS_5$ for $m_{n_a}=0$ is to be expected. Although $AdS_5$ can be embedded in $AdS_d$ for $d>5$ the embedding is such that the $AdS_5$ warped direction differs from that of $AdS_d$, as we discuss in Appendix~\ref{app:embedding}.

The constants $\beta_h^{(\vec{n})}$ are determined by the boundary conditions $\partial_z f_h^{(\vec{n})}|_{z_*}=0$, where $z_*=z_{0,L}=k^{-1}, e^{k\pi r_c}k^{-1}$, and the KK masses $m_{h,\vec{n}}$ follow from $\beta_{\vec{n}}(z_0)=\beta_{\vec{n}}(z_L)$, which to good approximation gives 
\begin{eqnarray}
J_{\nu_h-1}\left[\sqrt{m_{h,\vec{n}}^2-m_{n_a}^2}z_L\right]=0.
\end{eqnarray}
For $\sqrt{m_{h,\vec{n}}^2-m_{n_a}^2}\gg |(\nu_h-1)^2-1/4|e^{-k\pi r_c}k$ the KK masses may be approximated by:
\begin{eqnarray}
m_{h,\vec{n}}^2\simeq \pi^2(n+\frac{\nu_h}{2}-\frac{3}{4})^2 e^{-2k \pi r_c}k^2+m_{n_a}^2.
\end{eqnarray}
Considering the purely warped KK modes ($n_a=0$), one observes that as $d$ increases the mass of the KK modes increases whilst the relative KK spacing, $(m_{h,n+1}-m_{h,n})/m_{h,n}$, decreases. Thus the purely warped KK gravitons on a slice of $AdS_d$ are discernible from those of the RS model. The $AdS_d$ KK gravitons also differ from those found in spaces of the form $AdS_5\times \mathcal{M}^{d-5}$, given that purely warped gravitons in the latter match those of the RS model. The purely warped $AdS_d$ gravitons with $n>0$ couple to IR brane localized stress-energy sources with coupling $\Lambda_\pi^{-1}\gg M_{Pl}^{-1}$, as in the RS model. In particular one finds $\Lambda_\pi \simeq e^{-(d-3)k\pi r_c/2}M_{Pl}$, which is $\sim$~TeV for the parameters we consider. 
\section{Bulk Vectors on $\mathbf{AdS_{d}}$\label{sec:vec_ads_d}}
In this section we consider a bulk $U(1)$ gauge field in the $AdS_{d}$ background. As our ultimate purpose is to determine the viability of modeling a SM gauge boson by such a state the vector modes $M=\mu$ should have even parity to ensure a zero mode. The action for $A_M$ is
\begin{eqnarray}
S_{A}&=&-\frac{1}{4}\int d^{d}x\sqrt{G}\left\{G^{MP}G^{NQ}F_{MN}F_{PQ}\right\},
\end{eqnarray}
and we work with the conformal coordinates defined by (\ref{ads_conformal_metric}). The mixing between the vector mode and the gauge-scalar modes may be decoupled by introducing a bulk gauge fixing term,
\begin{eqnarray}
S_{GF}=-\frac{1}{2\xi}\int d^{d}x\frac{1}{(kz)^{d-4}}\left(\eta^{\nu\tau}\partial_\nu A_\tau+\xi (kz)^{d-4}[\sum_{\bar{a}}\partial_{\bar{a}}(KA_{\bar{a}})]\right)^2,
\end{eqnarray}
where we use the index $\bar{a}$ to denote $a,z$ so that $\sum_{\bar{a}}=\sum_{\bar{a}=a,z}$ and we define the quantity $K=K(z)$ by
\begin{eqnarray}
K\eta^{\nu\tau}=\sqrt{G}G^{\bar{a}\bar{a}}G^{\nu\tau}.
\end{eqnarray}
Varying the action $S_A+S_{GF}$ gives the bulk equations of motion,
\begin{eqnarray}
\sqrt{-G}G^{\mu\tau}G^{\nu\sigma}\partial_\mu F_{\tau\sigma}+\sum_{\bar{a}}\eta^{\mu\nu}\partial_{\bar{a}}[K\partial_{\bar{a}}A_\mu]+\frac{1}{\xi}\frac{1}{(kz)^{d-4}}\eta^{\mu\tau}\eta^{\nu\sigma}\partial_\mu\partial_\sigma A_\tau&=&0,\\
\eta^{\mu\tau}\partial_\tau\partial_\mu A_{a}+\xi \partial_{a} [(kz)^{d-4}\sum_{\bar{b}}\partial_{\bar{b}}(KA_{\bar{b}})]-\frac{1}{K}\sum_{\bar{b}}\partial_{\bar{b}}[\sqrt{G}G^{aa}G^{\bar{b}\bar{b}}F_{a\bar{b}}]&=&0\label{scalar_eom_1},\\
\eta^{\mu\tau}\partial_\tau\partial_\mu A_z+\xi \partial_{z} [(kz)^{d-4}\sum_{\bar{b}}\partial_{\bar{b}}(KA_{\bar{b}})]-\frac{1}{K}\sum_{a}\partial_{a}[\sqrt{G}G^{aa}G^{zz}F_{za}]&=&0\label{scalar_eom_2},
\end{eqnarray}
where the first equation describes the vector modes and the remaining $(d-4)$ equations are mixed and describe the gauge-scalars. Taking suitable combinations of (\ref{scalar_eom_1}) and (\ref{scalar_eom_2}) gives
\begin{eqnarray}
& &\eta^{\mu\tau}\partial_\tau\partial_\mu G_A-\frac{\xi}{K} \sum_{\bar{a}}\partial_{\bar{a}}\{ K\partial_{\bar{a}}G_A\}=0\label{scalar_eom_goldstone},\\
& &\eta^{\mu\tau}\partial_\tau\partial_\mu F_{za}+\sum_{\bar{b}}\partial_{z}\{\frac{1}{K}\partial_{\bar{b}}[\sqrt{G}G^{aa}G^{\bar{b}\bar{b}}F_{\bar{b}a}]\} +\partial_{a}\{\frac{1}{K}\sum_{b}\partial_{b}[\sqrt{G}G^{bb}G^{zz}F_{zb}]\}=0\label{gauge_scalar_eom},
\end{eqnarray}
where $G_A=(kz)^{d-4}\sum_{\bar{b}}\partial_{\bar{b}}(KA_{\bar{b}})$. Note that the states described by equation (\ref{scalar_eom_goldstone}) have decoupled and are, in fact, the Goldstone modes. The $(d-5)$ equations (\ref{gauge_scalar_eom}) remain mixed and describe the physical gauge-scalars.
\subsection{KK decomposition of the vector mode\label{sec:vector_kk}}
We expand the vector modes $A_\mu$ as
\begin{eqnarray}
A_\mu(x^\nu,x^a,z)=\sum_{\vec{n}}A^{(\vec{n})}_\mu(x^\nu)g^{(n_a)}_+(x^a)f^{(\vec{n})}_A(z),
\end{eqnarray}
and the profiles $f^{(\vec{n})}_A(z)$ must satisfy the following orthogonality relations:
\begin{eqnarray}
\int \frac{dz}{(kz)^{d-4}}f^{(n_a,n)}_Af^{(n_a,m)}_A&=&\delta^{mn},
\end{eqnarray}
and the equation of motion:
\begin{eqnarray}
\left[z^2\partial_z^2-(d-5)z\partial_z+m_{\vec{n}}^2z^2-m_{n_a}^2z^2\right]f^{(\vec{n})}_A=0.\label{vector_eom_f}
\end{eqnarray}
The solution to (\ref{vector_eom_f}) is
\begin{eqnarray}
f_A^{(\vec{n})}(z)=\frac{(kz)^{\nu_A}}{N_A^{(\vec{n})}}\left\{J_{\nu_A}\left[\sqrt{m_{\vec{n}}^2-m_{n_a}^2}z\right]+\beta_A^{(\vec{n})}Y_{\nu_A}\left[\sqrt{m_{\vec{n}}^2-m_{n_a}^2}z\right]\right\},\label{vector_profile}
\end{eqnarray}
where $N_A^{(\vec{n})}$ is a normalization constant and the order of the Bessel functions is
\begin{eqnarray}
\nu_A=\frac{1}{2}(d-3).
\end{eqnarray}
Equation (\ref{vector_profile}) generalizes the wavefunction for a bulk vector in the $AdS_5$ RS background to a higher dimensional slice of $AdS_{d}$. As such the $d=5$ case reduces to that of~\cite{Davoudiasl:1999tf}. The constants $\beta_A^{(\vec{n})}$ are determined by the boundary conditions $\partial_z f_A^{(\vec{n})}|_{z_*}=0$ and are found to be
\begin{eqnarray}
\beta^{(\vec{n})}_A(z_*)=-\frac{J_{\nu_A-1}\left[\sqrt{m_{\vec{n}}^2-m_{n_a}^2}z_*\right]}{Y_{\nu_A-1}\left[\sqrt{m_{\vec{n}}^2-m_{n_a}^2}z_*\right]},
\end{eqnarray}
with the KK masses $m_{\vec{n}}$ determined by solving $\beta^{(\vec{n})}_A(z_0)=\beta^{(\vec{n})}_A(z_L)$.  As with the KK gravitons, the wavefunction along the warped direction for a bulk vector differs from the RS result for $d>5$ with both the order of the Bessel functions and the power of the prefactor $(kz)$ increasing with $d$. The mass of the purely warped KK vectors ($n_a=0$) also increases with $d$ whilst the relative spacing of the KK modes $(m_{n+1}-m_{n})/m_{n}$ decreases. Thus the vector KK tower for $d>5$ is discernible from its RS counterpart. The KK action for the vector modes is finally given by
\begin{eqnarray}
\sum_{\vec{n}}\int d^4x\left\{-\frac{1}{4}\eta^{\mu\tau}\eta^{\nu\sigma}F^{(\vec{n})}_{\mu\nu}F^{(\vec{n})}_{\tau\sigma}-\frac{1}{2\xi}(\eta^{\nu\tau}\partial_\nu A^{(\vec{n})}_\tau)^2+\frac{1}{2}m_{\vec{n}}^2A^{(\vec{n})}A^{(\vec{n})}\right\},\label{ads_d_vector_kk_4D_action}
\end{eqnarray}
which reduces to the usual RS expression for $d=5$. We note that for $d>5$ the massless zero mode gauge boson has wave function
\begin{eqnarray}
f_A^{(0)}(z)=\sqrt{\frac{k(d-5)}{2}}\left[1-e^{-(d-5)k\pi r_c}\right]^{-1/2}\simeq\sqrt{\frac{k(d-5)}{2}},\label{vector_zero_profile}
\end{eqnarray}
which remains finite for $r_c\rightarrow \infty$ and differs from the $d=5$ case, for which $f_A^{(0)}(z)\propto r_c^{-1/2}$. This difference has been noted already in~\cite{Dubovsky:2000av}.
\subsection{Gauge-scalar modes in $\mathbf{AdS_6}$\label{sec:gauge_scalar_kk}}
For $d>5$ a bulk vector has $(d-5)$ additional degrees of freedom in the form of polarizations along the transverse directions. These, combined with the polarization in the warped direction, give rise to $(d-5)$ KK towers of physical gauge-scalars and a single KK tower of Goldstone modes; see e.g.~\cite{Burdman:2005sr} for studies of gauge-scalars in UED models. We shall not determine the KK towers of gauge-scalars for arbitrary $d$ in what follows but instead, as an example, provide the KK spectrum of scalar modes for $AdS_6$. The value $d=6$ is the smallest number of spacetime dimensions which admits a KK tower of physical gauge-scalars. For $AdS_6$ we take the transverse direction $x^5$ to be orbifolded as $S^1/Z_2'$, where the action of the orbifold symmetry is $Z_2': x^5\rightarrow- x^5$. The $(Z_2',Z_2)$ parities of a bulk $AdS_6$ gauge boson are
\begin{eqnarray}
A_\mu: (+,+)\quad,\quad A_5: (-,+)\quad,\quad A_z: (+,-).\label{ads_6_gauge_scalar_parity}
\end{eqnarray}
The parities for the scalar modes are fixed by the demand that $A_\mu$ be even and are such that $A_{5,z}$ do not posses zero modes. For $AdS_{6}$ the equations of motion (\ref{scalar_eom_goldstone}) and (\ref{gauge_scalar_eom}) reduce to
\begin{eqnarray}
\left\{z^2\partial^2_z-2z\partial_z-[\eta^{\mu\tau}\partial_\tau\partial_\mu -\partial_{5}^2]z^2\right\}G_A&=&0,\label{g_scalar_eom_goldstone2}\\
\left\{z^2\partial^2_z-2z\partial_z-[\eta^{\mu\tau}\partial_\tau\partial_\mu -\partial_{5}^2]z^2+2\right\}F_{5z}&=&0,\label{g_scalar_eom2}
\end{eqnarray}
where $G_A=(kz)^2\left[\partial_{5}(KA_{5})+\partial_{z}(KA_{z})\right]$. We KK expand the scalar modes as
\begin{eqnarray}
G_A(x^\mu,x^5,z)&=&\sum_{\vec{n}}m_{G,\vec{n}}A^{(\vec{n})}_{G}(x^\mu)g^{(n_5)}_+(x^5)f^{(\vec{n})}_G(z),\nonumber\\
F_{z5}(x^\mu,x^5,z)&=&\sum_{\vec{n}}m_{S,\vec{n}}A^{(\vec{n})}_{S}(x^\mu)g^{(n_5)}_-(x^5)f^{(\vec{n})}_S(z),
\end{eqnarray}
where $m_{G(S),\vec{n}}$ is the mass for the $\vec{n}$-th KK mode and $g^{(n_5)}_\pm$ are the usual even/odd parity wave functions for the $S^1/Z_2$ orbifold. The 4D fields satisfy
\begin{eqnarray}
\eta^{\mu\tau}\partial_\tau\partial_\mu A^{(\vec{n})}_{G}=-m_{G,\vec{n}}^2A^{(\vec{n})}_{G}\quad,\quad \eta^{\mu\tau}\partial_\tau\partial_\mu A^{(\vec{n})}_{S}=-m_{S,\vec{n}}^2A^{(\vec{n})}_{S},\label{gs_4d_field}
\end{eqnarray}
and the orthogonality relations are:
\begin{eqnarray}
\int \frac{dz}{(kz)^{2}}f^{(n_5,m)}_{S,G}f^{(n_5,n)}_{S,G}&=&\delta^{mn},\\
\int dx^5 g^{(m_5)}_\pm g^{(n_5)}_\pm&=&\delta^{m_5n_5}.
\end{eqnarray}
Using $\partial_{5}^2g^{(n_5)}_\pm=-m_{n_5}^2g^{(n_5)}_\pm$ and equation (\ref{gs_4d_field}) in the equations of motion gives:
\begin{eqnarray}
\left\{z^2\partial^2_z-2z\partial_z+(m_{G,\vec{n}}^2-m_{n_5}^2)z^2\right\}f^{(\vec{n})}_G&=&0,\label{g_scalar_eom_goldstone3}\\
\left\{z^2\partial^2_z-2z\partial_z+(m_{S,\vec{n}}^2-m_{n_5}^2)z^2+2\right\}f^{(\vec{n})}_S&=&0,\label{g_scalar_eom3}
\end{eqnarray}
which have solutions
\begin{eqnarray}
f_G^{(\vec{n})}(z)&=&\frac{(kz)^{3/2}}{N_G^{(\vec{n})}}\left\{J_{3/2}\left[\sqrt{m_{G,\vec{n}}^2-m_{n_a}^2}z\right]+\beta_G^{(\vec{n})}Y_{3/2}\left[\sqrt{m_{G,\vec{n}}^2-m_{n_a}^2}z\right]\right\},\label{goldstone_profile}\\
f_S^{(\vec{n})}(z)&=&\frac{(kz)^{3/2}}{N_S^{(\vec{n})}}\left\{J_{1/2}\left[\sqrt{m_{G,\vec{n}}^2-m_{n_a}^2}z\right]+\beta_S^{(\vec{n})}Y_{1/2}\left[\sqrt{m_{G,\vec{n}}^2-m_{n_a}^2}z\right]\right\}.\label{gauge_scalar_profile}
\end{eqnarray}
From (\ref{ads_6_gauge_scalar_parity}) one obtains the boundary conditions along the warped direction for $A_{5,z}$ as
\begin{eqnarray}
 A_5|=0\quad,\quad \partial_z A_z|=0,
\end{eqnarray}
which lead to
\begin{eqnarray}
\beta_S^{(\vec{n})}(z_*)=\beta_G^{(\vec{n})}(z_*)=-\frac{J_{1/2}\left[\sqrt{m_{S,\vec{n}}^2-m_{n_a}^2}z_*\right]}{Y_{1/2}\left[\sqrt{m_{S,\vec{n}}^2-m_{n_a}^2}z_*\right]},
\end{eqnarray}
and the KK masses follow from enforcing $\beta_S^{(\vec{n})}(z_0)=\beta_S^{(\vec{n})}(z_L)$.
Using the following KK expansions of $A_{5,z}$
\begin{eqnarray}
A_{5}(x^\mu,x^5,z)&=&\sum_{\vec{n}}A^{(\vec{n})}_{5}(x^\mu)g^{(n_5)}_+(x^5)f^{(\vec{n})}_5(z),\nonumber\\
A_{z}(x^\mu,x^5,z)&=&\sum_{\vec{n}}A^{(\vec{n})}_{z}(x^\mu)g^{(n_5)}_-(x^5)f^{(\vec{n})}_z(z),
\end{eqnarray}
one finds that $f_5^{(\vec{n})}(z)=f_G^{(\vec{n})}(z)$ and $f_z^{(\vec{n})}(z)=f_S^{(\vec{n})}(z)$, whilst the 4D fields are related as
\begin{eqnarray}
\frac{m_{S,\vec{n}}}{N_G^{(\vec{n})}}A^{(\vec{n})}_{G}&=&-\frac{m_{n_5}}{N_G^{(\vec{n})}}A^{(\vec{n})}_{5}+\frac{(m_{S,\vec{n}}^2-m_{n_5}^2)^{1/2}}{N_S^{(\vec{n})}}A^{(\vec{n})}_{z},\\
\frac{m_{S,\vec{n}}}{N_S^{(\vec{n})}}A^{(\vec{n})}_{S}&=&\frac{(m_{S,\vec{n}}^2-m_{n_5}^2)^{1/2}}{N_G^{(\vec{n})}}A^{(\vec{n})}_{5}+\frac{m_{n_5}}{N_S^{(\vec{n})}}A^{(\vec{n})}_{z}.
\end{eqnarray}
Combining the above gives the effective 4D action for the gauge-scalars,
\begin{eqnarray}
\sum_{\vec{n}}\frac{1}{2}\int d^4x\left\{\eta^{\mu\tau}\partial_\mu A_{G}^{(\vec{n})}\partial_\tau A_{G}^{(\vec{n})}- \xi m_{S,\vec{n}}^2(A^{(\vec{n})}_{G})^2+\eta^{\mu\tau}\partial_\mu A_{S}^{(\vec{n})}\partial_\tau A_{S}^{(\vec{n})}- m_{S,\vec{n}}^2(A^{(\vec{n})}_{S})^2\right\},\label{ads_6_gauge_scalar_4d_action}
\end{eqnarray}
and by adding (\ref{ads_6_gauge_scalar_4d_action}) to the $d=6$ case of the vector KK action (\ref{ads_d_vector_kk_4D_action}) one obtains the complete KK action for a bulk vector in $AdS_6$. Observe that in the unitary gauge $\xi\rightarrow \infty$ the modes $A^{(\vec{n})}_G$ become infinitely heavy and disappear from the spectrum so that, as advertised, these are the Goldstone modes which are `eaten' by the massive KK vectors $A^{(\vec{n})}_\mu$, $\vec{n}\ne0$. The modes $A_S^{(\vec{n})}$ are the physical gauge-scalars which remain in the spectrum in the unitary gauge.
\section{Bulk Fermions\label{sec:ads7_kk_fermi}}
In RS models the KK decomposition of a bulk 5D vectorial fermion produces a single chiral massless mode~\cite{Grossman:1999ra}. Being vectorial, an RS fermion may posses a bulk mass\footnote{Which must be odd under the $Z_2$ orbifold symmetry.} and by varying this mass over order one values (in units of $k$) the chiral mode is readily localized towards either the UV or IR brane~\cite{Grossman:1999ra,Gherghetta:2000qt}. By localizing the lighter (heavier) SM fermions towards the Planck (TeV) brane one may generate the observed SM fermion mass hierarchies with order one Yukawa couplings~\cite{Gherghetta:2000qt,rs_flavor} and thus the RS framework provides a mechanism by which to construct theories of flavor. We are interested in considering the generalization of the RS approach to flavor for $AdS_d$ with $d>5$. The vectorial nature of bulk RS fermions persists only in spacetimes with an odd number of dimensions; for even $d$ the minimal fermion is chiral with respect to the $d$-dimensional chiral projection operators and therefore does not admit a bulk mass. The RS approach to flavor is thus expected to generalize only for odd values of $d$ and we consider the simplest cases of odd $d>5$ in what follows and we obtain the KK spectrum for a bulk fermions on a slice of $AdS_7$ and $AdS_9$. We then consider the coupling of these bulk fermions, including the localizable chiral modes, to an IR brane scalar and a bulk vector in Section~\ref{sec:couplings}.

We point out that a bulk fermion on a slice of $AdS_7$ was considered already in~\cite{Appelquist:2002ft}. In that work the bulk fermion acquired an effective bulk mass by coupling to a bulk scalar with a non-vanishing background profile. The Yukawa coupling of the chiral zero-mode fermion to brane fields was then considered. Our analysis differs as we admit a mass for the bulk fermion and obtain the entire fermion KK spectrum; not just the zero mode profile as was done in~\cite{Appelquist:2002ft}. We then consider the Yukawa coupling of two such bulk fields to a brane scalar and the coupling of a bulk fermion to a bulk gauge boson. Our results and notation provide a transparent generalization of the familiar RS expressions.
\subsection{$\mathbf{AdS_7}$: Fermion orbifold parities\label{sec:ads7_grav}}
Before proceeding to discuss bulk $AdS_7$ fermions we specify the action of the orbifold symmetries acting in the extra dimensions. We write the index of the toroidal transverse dimensions as $a,b=5,6$, with the metric defined by
\begin{eqnarray}
ds^2_{AdS_7}&=&e^{-2\sigma(y)}\left[\eta_{\mu\nu}dx^\mu dx^\nu -\delta_{ab}dx^adx^b\right]-(dy)^{2},\nonumber\\
&\equiv&G_{MN}dx^Mdx^N\label{ads7_metric},
\end{eqnarray}
The extra dimensions $x^a,y$ are orbifolded via
\begin{eqnarray}
(T^2/Z_2')\times (S^1/Z_2),\label{ads7_obc}
\end{eqnarray}
with the action of $Z_2,'$, $Z_2$  defined by
\begin{eqnarray}
Z_2\ &:&\ y\rightarrow -y,\nonumber\\
Z_2'\ &:&\  x^a\rightarrow - x^a.
\end{eqnarray}
A bulk field in the above background is in general specified by two parities $(Z_2',Z_2)=(P',P)$, where $P',P=\pm $, and we note that the orbifolding (\ref{ads7_obc}) ensures there are no massless gravi-vectors in the spectrum. The action of the orbifold symmetries on a bulk fermion $\Psi$ is
\begin{eqnarray}
Z_2'&:&\Psi(x^\mu,x^a,y)\rightarrow \Psi'(x^\mu,-x^a,y)=i P' \Gamma^5\Gamma^6 \Psi(x^\mu,x^a,y),\\
Z_2&:&\Psi(x^\mu,x^a,y)\rightarrow \tilde{\Psi}(x^\mu,x^a,-y)=i P \Gamma^7 \Psi(x^\mu,x^a,y),
\end{eqnarray}
and our conventions for the 7D gamma matrices $\Gamma^M$ may be found in Appendix~\ref{app:ferm_not}, where we also discuss some general properties of 7D fermions. We shall work with $P=-1$ and $P'=+1$ so that the $(Z_2',Z_2)$ parities of the components of $\Psi$ are
\begin{eqnarray}
\Psi =\left(\begin{array}{c}\psi_{-R}\ (+,-)\\\psi_{-L}\ (-,-)\\\psi_{+L}\ (+,+)\\\psi_{+R}\ (-,+)\end{array}\right),\label{7d_fermi_parity}
\end{eqnarray}
and $\psi_{+L}$ is the only field which is even under both symmetries. Regardless of which values are used for $P',P$ there is always only one component of $\Psi$ which is even under both $Z_2'$ and $Z_2$; the selection of different values for $P',P$ simply determines which component is even. The action of the orbifold symmetries on a Dirac mass bilinear is
\begin{eqnarray}
Z_2'&:&\overline{\Psi}\Psi\rightarrow+\overline{\Psi}\Psi,\label{dirac_mass_obc}\\
Z_2&:&\overline{\Psi}\Psi\rightarrow-\overline{\Psi}\Psi,\label{dirac_mass_obc_1}
\end{eqnarray}
so that a bulk fermion may only have a Dirac mass if the mass is odd under the action of $Z_2$, as in the RS model.  
\subsection{$\mathbf{AdS_7}$: Fermion KK spectrum \label{sec:fermi_pro}}
The action for a bulk fermion in the $AdS_7$ background is:
\begin{eqnarray}
S_{\Psi}&=&\int d^7x\sqrt{G}\left\{\frac{i}{2}\overline{\Psi}\Gamma^{\underline{M}}e^M_{\underline{M}}\partial_M\Psi-\frac{i}{2}(\partial_M\overline{\Psi})\Gamma^{\underline{M}}e^M_{\underline{M}}\Psi-m_D\overline{\Psi}\Psi\right\},\label{7d_fermi_lagrangian}
\end{eqnarray}
where $e^M_{\underline{M}}=(kz)\delta^M_{\underline{M}}$. We have already dropped the spin connection terms, which arise from the use of the covariant derivative $D_M=\partial_M+\omega_M$, and cancel in the above. After rescaling the field $\Psi\rightarrow(kz)^3 \Psi$ and integrating by parts one has
\begin{eqnarray}
S_{\Psi}&=&\int d^7x\left\{i\overline{\Psi}\Gamma^{\mu}\partial_\mu\Psi+i\overline{\Psi}\Gamma^{7}\partial_7\Psi+i\overline{\Psi}\Gamma^{a}\partial_a\Psi-\frac{m_D}{kz}\overline{\Psi}\Psi\right\}.
\end{eqnarray}
We define the four component spinors $\psi_{+}=(\psi_{+ L},\psi_{+ R})^T$ and $\psi_{-}=(\psi_{- L},\psi_{- R})^T$ in terms of the component fields,
\begin{eqnarray}
\Psi_+=(0,0,\psi_{+L},\psi_{+R})^T\quad,\quad\Psi_-=(\psi_{-R},\psi_{-L},0,0)^T,
\end{eqnarray}
 and KK expand these four component fields as
\begin{eqnarray}
\psi_+(x^\mu,x^a,z)&=&\psi_{+ L}(x^\mu,x^a,z)+\psi_{+ R}(x^\mu,x^a,z)\nonumber\\
&=&\sum_{\vec{n}}\left\{\psi^{(\vec{n})}_{ L}(x^\mu)g^{(n_a)}_{+ L}(x^a)f^{(\vec{n})}_{+ L}(z) +\psi^{(\vec{n})}_{ R}(x^\mu)g^{(n_a)}_{+ R}(x^a)f^{(\vec{n})}_{+ R}(z)\right\},\nonumber\\
\psi_-(x^\mu,x^a,z)&=&\psi_{- L}(x^\mu,x^a,z)+\psi_{- R}(x^\mu,x^a,z)\nonumber\\
&=&\sum_{\vec{n}}\left\{\psi^{(\vec{n})}_{ L}(x^\mu)g^{(n_a)}_{- L}(x^a)f^{(\vec{n})}_{- L}(z) +\psi^{(\vec{n})}_{ R}(x^\mu)g^{(n_a)}_{- R}(x^a)f^{(\vec{n})}_{- R}(z)\right\},\nonumber
\end{eqnarray} 
where $\psi_{\pm L,R}=P_{L,R}\psi_\pm$. The wave functions obey the following orthogonality relations,
\begin{eqnarray}
\int dz (f^{*(n_a,m)}_{+L,R}f^{(n_a,n)}_{+ L,R}+f^{*(n_a,m)}_{-L,R}f^{(n_a,n)}_{-L,R})&=&\delta^{mn},\\
\int [\Pi_adx^a]g^{*(n_a)}_{\pm L}g^{(n_{b})}_{\pm L}=\int [\Pi_adx^a] g^{*(n_a)}_{\pm R}g^{(n_{b})}_{\pm R}&=&\delta^{n_an_{b}},
\end{eqnarray}
and the explicit form of the toroidal wave functions $g^{(n_a)}_{\pm L,R}$ is given in Appendix~\ref{app:t_wave_7}. With $m_{n_a}$ given by 
\begin{eqnarray}
m_{n_a}=\frac{\sqrt{n_5^2+n_6^2}}{R},\label{torus_kk_mass_7}
\end{eqnarray}
the equations of motion for the wave functions along the warped direction may be written as
\begin{eqnarray}
\left[\mp\partial_z-\frac{c}{z}\right]f^{(\vec{n})}_{\pm R}\pm m_{n_a}f^{(\vec{n})}_{\mp R}&=&-m_{\vec{n}}f^{(\vec{n})}_{\mp L},\label{ads7_vecfermi_eom_2a}\\
\left[\mp\partial_z-\frac{c}{z}\right]f^{(\vec{n})}_{\pm L}\pm m_{n_a}f^{(\vec{n})}_{\mp L}&=&-m_{\vec{n}}f^{(\vec{n})}_{\mp R},\label{ads7_vecfermi_eom_2b}
\end{eqnarray}
where $m_{\vec{n}}$ are the KK masses and the dimensionless mass $c$ is defined by $m_D=ck$.
The equations of motion  (\ref{ads7_vecfermi_eom_2a}), (\ref{ads7_vecfermi_eom_2b}) may be separated as
\begin{eqnarray}
(z^2 \partial_z^2 \mp c -c^2 +(m_{\vec{n}}^2-m_{n_a}^2)z^2)f^{(\vec{n})}_{\pm L,R}&=&0,
\end{eqnarray}
and, noting the parities (\ref{7d_fermi_parity}), one may use the equations of motion to obtain the boundary conditions,
\begin{eqnarray}
\left.f^{(\vec{n})}_{- L,R}\right|_{z_*}&=&0,\label{fermi_bc_odd}\\
\left.\left(\partial_z+\frac{c}{z}\right)f^{(\vec{n})}_{+L, R}\right|_{z_*}&=&0.\label{fermi_bc_even}
\end{eqnarray}
The solutions to the above are
\begin{eqnarray}
f_{\pm L,R}^{(\vec{n})}(z)&=&\frac{\sqrt{kz}}{N_{\pm \Psi}^{(\vec{n})}}\left\{J_{\nu_{\pm}}(\sqrt{m_{\vec{n}}^2-m_{n_a}^2}z)+\beta_{\Psi}^{(\vec{n})}Y_{\nu_{\pm}}(\sqrt{m_{\vec{n}}^2-m_{n_a}^2}z)\right\},\label{fermi_kk_+}
\end{eqnarray}
where the order of the Bessel functions is $\nu_{\pm}=|c\pm \frac{1}{2}|$ and the equations of motion require that the normalization constants satisfy
\begin{eqnarray}
N_{\pm \Psi}^{(\vec{n})} = \sqrt{\frac{2m_{\vec{n}}}{m_{\vec{n}}\pm m_{n_a}}}N_{\Psi}^{(\vec{n})}.
\end{eqnarray}
The KK masses are fixed by enforcing $\beta_{\Psi}^{(\vec{n})}(z_0)=\beta_{\Psi}^{(\vec{n})}(z_L)$, where
\begin{eqnarray}
\beta_{\Psi}^{(\vec{n})}(z_*)=-\frac{J_{\nu_{-}}(\sqrt{m_{\vec{n}}^2-m_{n_a}^2}z_*)}{Y_{\nu_{-}}(\sqrt{m_{\vec{n}}^2-m_{n_a}^2}z_*)},\label{fermi_beta}
\end{eqnarray}
and, similar to the RS case~\cite{Gherghetta:2000qt}, they may be approximated as
\begin{eqnarray}
m_{\vec{n}}^2 =m_{n,n_a}^2\simeq (n+\frac{c}{2}-\frac{1}{2})k^2\pi^2 e^{-2k\pi r_c}  +m_{n_a}^2,
\end{eqnarray}
for large $n$. Putting the above together the bulk fermion action reduces to the canonical KK form
\begin{eqnarray}
S_{\Psi}&=&\sum_{\vec{n}}\int d^4x\left\{i\bar{\psi}^{(\vec{n})}\gamma^{\mu}\partial_\mu\psi^{(\vec{n})}-m_{\vec{n}}\bar{\psi}^{(\vec{n})}\psi^{(\vec{n})}\right\}.\label{kk_fermi_lagrangian}
\end{eqnarray}

Our primary interest is in the spectrum of massless modes as this will determine the viability of employing a bulk 7D fermion. Consider first the case with $m_{n_a}=n_a=0$, for which equations (\ref{ads7_vecfermi_eom_2a}), (\ref{ads7_vecfermi_eom_2b}) have the solution $f^{(n,0)}_{\pm L,R}\propto z^{\mp c}$. However the boundary conditions (\ref{fermi_bc_odd}) force $f^{(n,0)}_{- L,R}=0$ and furthermore for $n_a=0$ one has $g^{(n_a=0)}_{+ R}=0$ as $\psi_{+R}$. Thus the only non-vanishing mode is $f^{(0,0)}_{+ L}\propto z^{-c}$ with the normalized wavefunction
\begin{eqnarray}
(kz)^3f_{+L}^{(0,0)}(z)=\sqrt{\frac{k(1/2-c)}{(kz_L)^{(1-2c)}-1}}(kz)^{3-c},\label{ads7_fermi_nomass}
\end{eqnarray}
where for completeness we retain the factor of $(kz)^3$ previously scaled out. This is identical to the usual RS profile~\cite{Grossman:1999ra,Gherghetta:2000qt} modulo the replacement $(kz)^2\rightarrow (kz)^3$ for the factor scaled out in the above decomposition. One can easily show that no massless modes obtain when $m_{n_a}\ne0$ so the chiral mode (\ref{ads7_fermi_nomass}) is the only massless mode in the spectrum.
\subsection{$\mathbf{AdS_9}$: Fermion orbifold parities}
As it will be helpful in what follows to be able to distinguish between the different transverse directions we use a slightly different notation for the transverse coordinate labels in this section and write the index of the toroidal dimensions as $a,a'=5,6$ and $b,b'=7,8$ with the metric defined by
\begin{eqnarray}
ds^2_{AdS_9}&=&e^{-2\sigma(y)}\left[\eta_{\mu\nu}dx^\mu dx^\nu -\delta_{aa'}dx^adx^{a'}-\delta_{bb'}dx^bdx^{b'}\right]-(dy)^{2},\nonumber\\
&\equiv&G_{MN}dx^Mdx^N\label{ads9_metric},
\end{eqnarray}
The extra dimensions $x^{a,b},y $ are orbifolded via
\begin{eqnarray}
(T^2/Z_2')\times (T^2/Z_2'')\times(S^1/Z_2),\label{ads9_obc}
\end{eqnarray}
with the action of the orbifold symmetry defined by (\ref{ads7_obc}) and
\begin{eqnarray}
Z_2''\ :\  x^b\rightarrow - x^b,
\end{eqnarray}
The action of the orbifold symmetries on a bulk fermion $\Psi$ is 
\begin{eqnarray}
Z_2&:&\Psi(x^\mu,x^a,x^b,y)\rightarrow \tilde{\Psi}(x^\mu,x^a,x^b,-y)=i P \g^7 \Psi(x^\mu,x^a,x^b,y),\\
Z_2'&:&\Psi(x^\mu,x^a,x^b,y)\rightarrow \Psi'(x^\mu,-x^a,x^b,y)=i P' \g^5\g^6 \Psi(x^\mu,x^a,x^b,y),\\
Z_2''&:&\Psi(x^\mu,x^a,x^b,y)\rightarrow \Psi'(x^\mu,x^a,-x^b,y)=i P'' \g^7\g^8 \Psi(x^\mu,x^a,x^b,y),
\end{eqnarray}
where the three parities $P,P',P''$ all take the values $\pm1$. Here $\g^M$ the 9D Dirac gamma matrices; our conventions for which may be found in Appendix~\ref{app:ferm_not_9}, where we also discuss some general properties of 9D fermions. We shall work with $P=-1$, $P'=+1$ and $P''=-1$ so the $(Z_2,Z_2'',Z_2')$ parities for the components of $\Psi =( \psi_\downarrow,\psi_\uparrow)^T$ are
\begin{eqnarray}
\Psi_{\downarrow} =\left(\begin{array}{c}\psi_{1L}\ (-,-,+)\\\psi_{1R}\ (-,-,-)\\\psi_{2R}\ (-,+,+)\\\psi_{2L}\ (-,+,-)\end{array}\right)\quad,\quad\Psi_{\uparrow} =\left(\begin{array}{c}\psi_{3R}\ (+,-,+)\\\psi_{3L}\ (+,-,-)\\\psi_{4L}\ (+,+,+)\\\psi_{4R}\ (+,+,-)\end{array}\right),\label{9d_fermi_parity}
\end{eqnarray}
and $\psi_{4L}$ is the only completely even field. Regardless of which values are used for the parities $P,P',P''$ only one component of $\Psi$ is even under $Z_2$, $Z_2'$ and $Z_2''$; the selection of different parities simply determines which component is completely even. The action of the orbifold symmetries $Z_2',Z_2$ on a Dirac mass bilinear is again given by equations (\ref{dirac_mass_obc}) and (\ref{dirac_mass_obc_1}) whilst the action of $Z_2''$ is
\begin{eqnarray}
Z_2''&:&\overline{\Psi}\Psi\rightarrow+\overline{\Psi}\Psi.
\end{eqnarray}
\subsection{$\mathbf{AdS_9}$: Fermion KK spectrum \label{sec:fermi_pro9}}
The action for a bulk fermion in the $AdS_9$  background is
\begin{eqnarray}
S_{\Psi}&=&\int d^9x\sqrt{G}\left\{\frac{i}{2}\overline{\Psi}\g^{\underline{M}}e^M_{\underline{M}}\partial_M\Psi-\frac{i}{2}(\partial_M\overline{\Psi})\g^{\underline{M}}e^M_{\underline{M}}\Psi-m_D\overline{\Psi}\Psi\right\},\label{9d_fermi_lagrangian}
\end{eqnarray}
where $e^M_{\underline{M}}=(kz)\delta^M_{\underline{M}}$ and we have already dropped the spin connection terms which cancel in the above. After rescaling the field $\Psi\rightarrow(kz)^4 \Psi$ and integrating by parts one has
\begin{eqnarray}
S_{\Psi}&=&\int d^9x\left\{i\overline{\Psi}\g^{M}\partial_M\Psi-\frac{m_D}{kz}\overline{\Psi}\Psi\right\}.
\end{eqnarray}
We define the four component spinors $\psi_{\alpha}=(\psi_{\alpha L},\psi_{\alpha R})^T$ with $\alpha=1,2,3,4$, in terms of the component fields,
\begin{eqnarray}
\Psi=(\psi_{1 L},\psi_{1R},\psi_{2R},\psi_{2L},\psi_{3R},\psi_{3L},\psi_{4L},\psi_{4R})^T,\label{9d_fermi_compon_label}
\end{eqnarray}
and the KK expansion for the four component fermions is
\begin{eqnarray}
\psi_\alpha(x^\mu,x^{a,b},z)&=&\psi_{\alpha L}(x^\mu,x^{a,b},z)+\psi_{\alpha R}(x^\mu,x^{a,b},z)\nonumber\\
&=&\sum_{\vec{n}}\left\{\psi^{(\vec{n})}_{ L}(x^\mu)g^{(n_a)}_{\alpha L}(x^a) h^{(n_b)}_{\alpha L}(x^b)f^{(\vec{n})}_{\alpha L}(z)+\psi^{(\vec{n})}_{ R}(x^\mu)g^{(n_a)}_{\alpha R}(x^a)h^{(n_b)}_{\alpha R}(x^b)f^{(\vec{n})}_{\alpha R}(z)\right\}.\nonumber
\end{eqnarray} 
The wave functions obey the following orthogonality relations,
\begin{eqnarray}
\sum_{\alpha=1}^4\int dz f^{*(m,n_a,n_b)}_{\alpha L}f^{(n,n_a,n_b)}_{\alpha L}&=&\delta^{mn},\\
\int [\Pi_adx^a] g^{*(n_a)}_{\alpha L}g^{(n_{a'})}_{\alpha L}&=&\delta^{n_an_{a'}},\\
\int [\Pi_bdx^b] h^{*(n_b)}_{\alpha L}h^{(n_{b'})}_{\alpha L}&=&\delta^{n_bn_{b'}},
\end{eqnarray}
and similarly with the replacement $L\rightarrow R$. The explicit form of $g^{(n_a)}_{\alpha L,R}$, $h^{(n_b)}_{\alpha L,R}$ are given in Appendix~\ref{t_wave_9}. In terms of the masses
\begin{eqnarray}
m_{n_a}=\frac{\sqrt{n_5^2+n_6^2}}{R}\quad,\quad m_{n_b}=\frac{\sqrt{n_7^2+n_8^2}}{R},\label{torus_kk_mass_9}
\end{eqnarray}
the equations of motion for the warped direction wave functions are
\begin{eqnarray}
\left[-\partial_z-\frac{c}{z}\right]f^{(\vec{n})}_{3 R}-m_{n_b}f^{(\vec{n})}_{2 R}-m_{n_a}f^{(\vec{n})}_{1 R}&=&-m_{\vec{n}}f^{(\vec{n})}_{1 L},\label{ads9_vecfermi_eom_2a}\\
\left[-\partial_z-\frac{c}{z}\right]f^{(\vec{n})}_{4 R}-m_{n_b}f^{(\vec{n})}_{1 R}+m_{n_a}f^{(\vec{n})}_{2 R}&=&-m_{\vec{n}}f^{(\vec{n})}_{2 L},\label{ads9_vecfermi_eom_2b}\\
\left[\partial_z-\frac{c}{z}\right]f^{(\vec{n})}_{1 R}+m_{n_b}f^{(\vec{n})}_{4 R}+m_{n_a}f^{(\vec{n})}_{3 R}&=&-m_{\vec{n}}f^{(\vec{n})}_{3 L},\label{ads9_vecfermi_eom_2c}\\
\left[\partial_z-\frac{c}{z}\right]f^{(\vec{n})}_{2 R}+m_{n_b}f^{(\vec{n})}_{3 R}-m_{n_a}f^{(\vec{n})}_{4 R}&=&-m_{\vec{n}}f^{(\vec{n})}_{4 L},\label{ads9_vecfermi_eom_2d}
\end{eqnarray}
where $m_{\vec{n}}$ are the KK masses and the dimensionless mass $c$ is again defined by $m_D=ck$. The wave functions $f_{\alpha L,R}^{(\vec{n})}$ must also satisfy the four equations obtained by replacing $f_{\alpha L}^{(\vec{n})}\leftrightarrow f_{\alpha R}^{(\vec{n})}$ in (\ref{ads9_vecfermi_eom_2a})-(\ref{ads9_vecfermi_eom_2d}). Noting the parities (\ref{9d_fermi_parity}) one may use the equations of motion to obtain the boundary conditions,
\begin{eqnarray}
\left.f^{(\vec{n})}_{\alpha L,R}\right|_{z_*}&=&0\quad \mathrm{for}\quad\alpha=1,2,\label{fermi_bc_odd_9}\\
\left.\left(\partial_z+\frac{c}{z}\right)f^{(\vec{n})}_{\alpha L, R}\right|_{z_*}&=&0\quad \mathrm{for}\quad\alpha=3,4.\label{fermi_bc_even_9}
\end{eqnarray}
Equations (\ref{ads9_vecfermi_eom_2a})-(\ref{ads9_vecfermi_eom_2d}) may be separated as
\begin{eqnarray}
(z^2 \partial_z^2 + c -c^2 +\tilde{m}_{\vec{n}}^2z^2)f^{(\vec{n})}_{\alpha L,R}&=&0\quad \mathrm{for}\quad\alpha=1,2,\\
(z^2 \partial_z^2 - c -c^2 +\tilde{m}_{\vec{n}}^2z^2)f^{(\vec{n})}_{\alpha L,R}&=&0\quad \mathrm{for}\quad\alpha=3,4,
\end{eqnarray}
where we define $\tilde{m}_{\vec{n}}^2\equiv m_{\vec{n}}^2-m_{n_a}^2-m_{n_b}^2$. The solutions are,
\begin{eqnarray}
f_{\alpha L,R}^{(\vec{n})}(z)&=&\frac{\sqrt{kz}}{N_{\alpha L,R}^{(\vec{n})}}\left\{J_{\nu_{-}}(\tilde{m}_{\vec{n}}z)+\beta_{\Psi}^{(\vec{n})}Y_{\nu_{-}}(\tilde{m}_{\vec{n}}z)\right\}\quad \mathrm{for}\quad\alpha=1,2,\label{fermi_kk_alp}\\
f_{\alpha L,R}^{(\vec{n})}(z)&=&\frac{\sqrt{kz}}{N_{\alpha L,R}^{(\vec{n})}}\left\{J_{\nu_{+}}(\tilde{m}_{\vec{n}}z)+\beta_{\Psi}^{(\vec{n})}Y_{\nu_{+}}(\tilde{m}_{\vec{n}}z)\right\}\quad \mathrm{for}\quad\alpha=3,4,\label{fermi_kk_bet}
\end{eqnarray}
where the order of the Bessel functions is $\nu_{\pm}=|c\pm \frac{1}{2}|$ and we have used of the equations of motion. The normalization constants are not independent and may be expressed in terms of a single constant, as given in Appendix~\ref {app:norm}. The KK masses $m_{\vec{n}}$ are found by solving $\beta_{\Psi}^{(\vec{n})}(z_0)=\beta_{\Psi}^{(\vec{n})}(z_L)$, with the constants $\beta_{\Psi}^{(\vec{n})}(z_*)$ given by 
\begin{eqnarray}
\beta_{\Psi}^{(\vec{n})}(z_*)=-\frac{J_{\nu_{-}}(\sqrt{m_{\vec{n}}^2-m_{n_a}^2-m_{n_b}^2}z_*)}{Y_{\nu_{-}}(\sqrt{m_{\vec{n}}^2-m_{n_a}^2-m_{n_b}^2}z_*)}.\label{fermi_beta_9}
\end{eqnarray}
The spectrum contains a single chiral massless mode with profile
 \begin{eqnarray}
(kz)^4f_{4L}^{(0)}(z)=\sqrt{\frac{k(1/2-c)}{(kz_L)^{(1-2c)}-1}}(kz)^{4-c},\label{ads9_fermi_nomass}
\end{eqnarray}
which matches the RS result modulo the replacement $(kz)^2\rightarrow (kz)^4$ for the scale factor. For large $n$ the KK masses may be approximated as~\cite{Gherghetta:2000qt}
\begin{eqnarray}
m_{\vec{n}}^2 \simeq (n+\frac{c}{2}-\frac{1}{2})k^2\pi^2 e^{-2k\pi r_c}  +m_{n_a}^2+m_{n_b}^2,
\end{eqnarray}
and putting the above together the bulk fermion action reduces to the canonical KK form. 

Before proceeding to consider the coupling of bulk fermions in $AdS_{7,9}$ to bosons we note that, relative to the RS model, the order of the Bessel functions in the fermion profiles (\ref{fermi_kk_alp}), (\ref{fermi_kk_bet}) has not changed as a result of having increased $d$ from the RS value of $d=5$ to $d=7,9$. This differs from the explicit $d$ dependence found earlier for bulk vectors and gravitons. As noted already for the zero modes, the power of the factor initially scaled out of the fermion wave functions does increase with $d$ so the fermion profiles do display some $d$ dependence.
\section{Coupling to a Brane Scalar and a Bulk Gauge Field\label{sec:couplings}}
We have seen that localizable chiral zero mode fermions, familiar from RS models, may also be obtained in \ads and \ad. In this section we consider an IR brane scalar to show that the RS warped realization of the weak scale also carries over to $AdS_d$ and, by coupling two bulk fermions to such a brane scalar, we show that for $AdS_{7,9}$ the RS approach to flavour also generalizes. We then consider the coupling of a bulk fermion to a bulk gauge field for $AdS_{7,9}$ and show that, in the phenomenologically interesting case of $R^{-1}\sim$~TeV, the effective 4D coupling for the zero modes experiences volume suppression. After these considerations we shall comment on the model building possibilities in $AdS_d$ and contrast these with spaces where the transverse directions are external to the warping, $AdS_5\times \mathcal{M}^{d-5}$, with an emphasis on the $AdS_5\times T^2$ case~\cite{McDonald:2009md}.

Consider a $(d-1)$ dimensional scalar $\Phi$ localized on the IR brane of a slice of $AdS_d$ with the usual quartic potential:
\begin{eqnarray}
S_\Phi&=&\int d^{d}x\sqrt{\bar{G}}\left\{G^{\bar{M}\bar{N}}\partial_{\bar{M}}\Phi^\dagger \partial_{\bar{N}}\Phi -\frac{\lambda}{M_*^{d-5}}(\Phi^2-v_0^2M_*^{d-5})^2\right\}\delta(y-\pi r_c),\nonumber\\
&=&\int d^{d-1}x\left\{\eta^{\bar{M}\bar{N}}\partial_{\bar{M}}\Phi^\dagger \partial_{\bar{N}}\Phi -\frac{\lambda}{e^{(5-d)k\pi r_c}M_*^{d-5}}\left(\Phi^2-\frac{v_0^2M_*^{d-5}}{e^{(d-3)k\pi r_c}}\right)^2\right\},
\end{eqnarray}
where $[\Phi]=(d-3)/2$ and the VEV is written in terms of the dimension one parameter $[v_0]=1$. The barred quantities denote the restriction to the brane at $y=\pi r_c$ and we have rescaled $\Phi\rightarrow e^{(d-3)k\pi r_c/2}\Phi$. The vacuum value of $\Phi$ is $\langle \Phi^{(0)}\rangle=v_0M_*^{(d-5)/2}e^{(3-d)k\pi r_c/2}$ and the natural scale for $v_0$ is $v_0\sim k$. Noting that the zero mode has the wavefunction $\Phi^{(0)}=\phi^{(0)}(x)/(2\pi R)^{(d-5)/2}$ the VEV for the 4D field is
\begin{eqnarray}
\langle \phi^{(0)}\rangle&=&v\equiv \frac{v_0}{e^{k\pi r_c}}\left[\frac{M_*2\pi R}{e^{k\pi r_c}}\right]^{\frac{d-5}{2}},\label{higgs_vev_ads}
\end{eqnarray}
where, in the case of electroweak symmetry breaking, $v\sim 246$~GeV would be the electroweak scale. The $d=5$ case of (\ref{higgs_vev_ads}) reproduces the usual RS expression for the VEV of an IR brane scalar whilst for $d>5$ it generalizes the RS expression to $AdS_d$. We are working with the phenomenologically interesting case of $R^{-1}\sim$~TeV and as $e^{-k\pi r_c}M_*\sim$~TeV the factor in brackets in (\ref{higgs_vev_ads}) is $\mathcal{O}(1-10)$. In particular for $e^{-k\pi r_c}M_*2\pi R\sim 1$ the weak scale is $v\sim e^{-k\pi r_c}v_0$ so that, as in the RS model, the weak scale is realized via the warped suppression of the order $\sim k$ input parameter\footnote{We note that equation (\ref{higgs_vev_ads}) seems to indicate that for $RM_*\sim\mathcal{O}(1)$ the 4D Higgs VEV is $\langle \phi^{(0)}\rangle\sim v_0e^{-(d-3)k\pi r_c/2}$. However, as we show in Section~\ref{sec:eff_4d_theory}, the effective 4D quartic coupling for the IR brane Higgs becomes non-perturbative in this region of parameter space so it is not clear that this deduction can be trusted.} $v_0$.

The RS approach to flavor also carries through to $AdS_7$ and $AdS_9$, with the Yukawa Lagrangian between two bulk fermions $\Psi_{1,2}$ and an IR brane scalar being
 \begin{eqnarray}
S_{Yuk}&=&-\frac{\lambda_Y}{M_*^{(d-3)/2}} \int d^{d}x \sqrt{\bar{G}}\Phi\Psi_1\Psi_2\delta(z-z_L)\nonumber\\
&=&-\int d^4x\psi^{(0)}_1\psi^{(0)}_2  \left[m_{12}+\lambda_y\phi^{(0)}\right]+ ....\label{yuk_rs7}
\end{eqnarray}
where the dots denote terms containing modes with $\vec{n}>0$ and  the fermion mass is $m_{12}=\lambda_y v$, with the effective 4D Yukawa coupling between the zero modes defined as\footnote{The numerical subscripts here label the different fermion fields $\Psi_{1,2}$ and not different spinor components.}  
\begin{eqnarray}
\lambda_y = \frac{\lambda_Y}{e^{-k \pi r_c}M_*}\left[\frac{e^{k\pi r_c}}{{M_*2\pi R}}\right]^{\frac{d-5}{2}}f_1^{(0)}(z_L)f_2^{(0)}(z_L)\quad,\quad d=5,7,9.
\end{eqnarray}
For $d=5$ this reproduces the familiar expression for the effective 4D Yukawa coupling in RS models~\cite{Grossman:1999ra,Gherghetta:2000qt} whilst the $d=7,9$ cases generalize the RS result and show that the RS approach to flavour holds for the warped spaces \ads and \ad.

We may also consider the coupling between a bulk fermion and a bulk gauge boson in $AdS_{7,9}$:
\begin{eqnarray}
S_{\Psi,A}&=&\frac{g_d}{M_*^{(d-4)/2}}\int d^dx\sqrt{G}e^M_{\underline{M}}\overline{\Psi}\Gamma^{\underline{M}}\Psi A_M\nonumber\\
&=&g_4\int d^4x\bar{\psi}_L^{(0)}\gamma^{\mu}\psi^{(0)}_L A_\mu^{(0)}+.... ,\label{7d_fermi_gauge}
\end{eqnarray}
where $g_d$ is a dimensionless bulk gauge coupling and for $d=7$ ($d=9$) the gamma matrices are the 7D (9D) Dirac matrices given in Appendix~\ref{app:ferm_not} (Appendix~\ref{app:ferm_not_9}). In the last line we have retained only the terms with the chiral mode and defined the 4D gauge coupling as
\begin{eqnarray}
 g_4&=&\frac{g_d}{M_*^{(d-4)/2}}\frac{1}{(2\pi R)^{(d-5)/2}}\int dz f^{*(0)}_{+L}f_{+L}^{(0)}f_A^{(0)}.\label{RS_like_coupling_ffa}
\end{eqnarray}
Using the vector zero mode profile for $d>5$ (\ref{vector_zero_profile}) gives
\begin{eqnarray}
 g_4&\simeq&g_dM_*\left[\frac{2k(d-5)}{M_*^{(d-2)}(2\pi R)^{(d-5)}}\right]^{1/2}\nonumber\\
&\sim&g_d\frac{M_*}{M_{Pl}},\label{RS_like_coupling_ffa2x}
\end{eqnarray}
where we have used the leading order expression for the 4D Planck mass via (\ref{4d_planck}). One readily observes a volume suppression of the effective 4D couplings. For example, with $k\sim M_*$ and $R^{-1}\sim1$~TeV one has 
\begin{eqnarray}
e^{k\pi r_c}\simeq \frac{M_*}{\mathrm{TeV}}\simeq \left[\frac{d-3}{2 (2\pi)^{d-5}}\right]^{\frac{1}{d-3}}\times \left[\frac{M_{Pl}}{\mathrm{TeV}}\right]^{\frac{2}{d-3}},
\end{eqnarray}
and provided $d$ is not too large this gives $M_{Pl}\sim e^{(d-3)k\pi r_c/2}~\mathrm{TeV}$ so that
\begin{eqnarray}
 g_4\sim g_d e^{(5-d)k\pi r_c/2}= g_d\times\left\{\begin{array}{ccccc}e^{-k\pi r_c}&\sim&10^{-15/2}&\mathrm{for}&AdS_7\\e^{-2k\pi r_c} &\sim& 10^{-10}&\mathrm{for}&AdS_9\end{array}\right.\label{RS_like_coupling_ffa3x}.
\end{eqnarray}

From the above considerations we may surmise the following. The RS realization of the weak scale via spacetime warping carries over to a slice of $AdS_d$ for $d>5$ when the Higgs boson is a $(d-1)$ dimensional field localized on the IR brane. The RS approach to flavour also carries over to \ads and $AdS_9$, with the coupling between two chiral zero mode fermions and a brane localized scalar sensitive to the fermion bulk mass parameters such that hierarchical Yukawas are expected in the 4D theory. However for the phenomenologically interesting case of $R^{-1}\sim1$~TeV the effective 4D gauge coupling between a chiral zero mode fermion and the zero mode of a bulk gauge field experiences volume suppression. The severity of this suppression increases with $d$, though already for $d=7$ it is of order $\sim 10^{-15/2}$. A similar volume suppression is known to occur for models with large extra dimensions~\cite{ArkaniHamed:1998nn}.

Consequently bulk SM fermions and gauge fields are not viable for both \ads and $AdS_9$. The appealing RS approach to flavour is therefore successful only on a slice of $AdS_5$ or, as shown in~\cite{McDonald:2009md} for the case of $AdS_5\times T^2$, for certain spaces of the form $AdS_5\times \mathcal{M}^{d-5}$. If SM matter fields propagate in the transverse dimensions of $AdS_d$ one expects them to be localized at either the UV or IR brane, with the weak scale realized via warping in the latter case. Actually, as we shall show in Section~\ref{sec:eff_4d_theory}, for $R^{-1}\sim$~TeV the effective 4D couplings are highly suppressed for UV localization so that only IR brane localization of the SM is viable.

Note that fermions which are sterile with respect to the SM gauge group may propagate in the bulk, with such a scenario considered already for \ads in the context of a brane localized UED model in~\cite{Appelquist:2002ft}. We further note that extended gauge sectors can also propagate in the bulk and the gauge coupling volume suppression can motivate a very weak coupling for such sectors. As an example consider the localization of the SM on the IR brane of $AdS_6$. This would realize an embedding of the minimal UED model on the IR brane and simultaneously motivate the weak/Planck hierarchy. With the gauge group extension $\mathcal{G}_{SM}\times \mathcal{G}_X$, the $\mathcal{G}_X$-symmetry could be broken on the IR brane at a scale of $\sim$~TeV and yet remain experimentally viable if it propagates in the bulk. The effective couplings in the 4D theory would be of order $g_6 M_*/M_{Pl}\sim g_6 e^{-k\pi r_c/2}\sim 10^{-5}g_6$ and are therefore automatically suppressed. Such a scenario may offer an interesting way to employ, for example, a weakly coupled symmetry which plays a custodial role and is broken at the weak scale. 

The main model building feature of the $AdS_d$ spaces seems to be the ability to combine the warped explanation for the weak/Planck hierarchy with the KK parity found in UED models. In UED models KK parity is a residual from an underlying spacetime isometry. The transverse space in $AdS_d$ admits such an isometry so that KK parity may remain viable when the SM fields propagate in the transverse space. In particular if the SM fields are localized on the $(d-1)$ dimensional IR brane of $AdS_d$ one obtains a geometrical motivation for both the weak/Planck hierarchy and the existence of stable dark matter. The main experimental signature for the $AdS_d$ spaces in this instance would be the observation of warped KK gravitons in addition to the UED KK modes. Such a signature also occurs when the $(d-1)$ dimensional UED model is realized by embedding the SM fields on the IR brane of $AdS_5\times T^{d-5}$, as discussed in~\cite{McDonald:2009md}. However the graviton KK towers on $AdS_d$ differ from $AdS_5\times T^{d-5}$ so that if a $(d-1)$ dimensional UED scenario is discovered one would be able to experimentally determine if the UED model is embedded in either of these distinct warped spaces by carefully studying the graviton KK spectrum.

We note that for $RM_*\sim\mathcal{O}(1)$ one has $M_*\sim M_{Pl}$ and the volume suppression observed in (\ref{RS_like_coupling_ffa2x}) disappears. Although the transverse KK modes disappear from the low energy spectrum in this limit, this case may be interesting unto itself. However, as we show in the next section, it is not clear at present what the correct description of the IR brane (including the localized Yukawa coupling) should be in this instance. It should also be stated that whilst the transverse KK (or dark matter) scale in $AdS_d$ is set by $R^{-1}$, there is no \emph{a priori} connection between this scale and the weak scale. Thus the usual WIMP paradigm requires the transverse radius to be stabilized at $R^{-1}\sim$~TeV~$\ll M_*$. As we will show below, such a condition is in any case necessary for the validity of the effective theory description we have employed. This situation is to be contrasted with $AdS_5\times T^2$~\cite{McDonald:2009md} where the underlying geometry also motivates the weak/Planck hierarchy (via warping) and dark matter (via KK parity as an isometry remnant). In that case the transverse KK scale is automatically warped to the IR brane scale, so that once the weak/Planck hierarchy is established via warping the dark matter scale is also $\sim$~TeV, even if the transverse radius is stabilized at $R^{-1}\sim M_*$.
\section{Validity of the Effective 4D Description\label{sec:eff_4d_theory}}
Throughout the present work we have assumed a transverse compactification scale of $R^{-1}\sim$~TeV. There are two reasons for having restricted our attention to this case. The first reason is phenomenological as the new KK modes associated with the transverse space in $AdS_d$ will be accessible to colliders only for $R^{-1}$ of order TeV and the lightest transverse KK mode may also be a good DM candidate for a TeV scale compactification. The second reason is theoretical as the effective 4D theory description on the IR brane breaks down for $R^{-1}>$~TeV when the IR brane scale is $\sim$~TeV. We briefly demonstrate the latter point in what follows. To this end we use $AdS_7$ as an example and consider a non-interacting 6D scalar field localized on the IR brane:
\begin{eqnarray}
S_\Phi&=&\frac{1}{2}\int d^{7}x\sqrt{\bar{G}}\left\{G^{\bar{M}\bar{N}}\partial_{\bar{M}}\Phi \partial_{\bar{N}}\Phi -m_{\Phi}^2\Phi^2\right\}\delta(y-\pi r_c),\nonumber\\
&=&\frac{1}{2}\int d^{6}xe^{-4k\pi r_c}\left\{\eta^{\bar{M}\bar{N}}\partial_{\bar{M}}\Phi \partial_{\bar{N}}\Phi -\frac{m_{\Phi}^2}{e^{2k\pi r_c}}\Phi^2\right\},\nonumber\\
&=&\frac{1}{2}\int d^{6}x\left\{\eta^{\bar{M}\bar{N}}\partial_{\bar{M}}\Phi \partial_{\bar{N}}\Phi -\frac{m_{\Phi}^2}{e^{2k\pi r_c}}\Phi^2\right\},
\end{eqnarray}
where the barred quantities denote brane restriction. To obtain the last line we have rescaled the field $\Phi\rightarrow e^{2k\pi r_c}\Phi$ to bring the kinetic term in the $x^\mu$ directions into a canonical form. With the KK expansion
\begin{eqnarray}
\Phi(x^\mu,x^a)=\sum _{n_a}\phi^{(n_a)}(x^\mu)g^{(n_a)}(x^a),
\end{eqnarray}
where the profiles obey $\partial_a^2 g^{(n_a)}= -m_{n_a}^2g^{(n_a)}$ with $m_{n_a}\sim R^{-1}$, the action reduces to the standard KK form:
\begin{eqnarray}
S_\Phi&=&\frac{1}{2}\sum_{n_a}\int d^{4}x\left\{\eta^{\mu\nu}\partial_{\mu}\phi^{(n_a)} \partial_{\nu}\phi^{(n_a)} -m_{\phi,n_a}^2(\phi^{(n_a)})^2\right\}.
\end{eqnarray}
The KK masses are
\begin{eqnarray}
m_{\phi,n_a}^2= m_{\Phi}^2e^{-2k\pi r_c}+m_{n_a}^2,\label{scalar_kk}
\end{eqnarray}
where the bare mass is warped down as $m_\Phi e^{-k\pi r_c}$ whilst the KK mass $m_{n_a}$ is not. As the effective transverse radius in the 4D theory is not warped it may lie below the cutoff of the 7D theory and yet exceed the warped down cutoff on the IR brane, that is $R$ may lie in the range $M_*>R^{-1}>\Lambda_{IR}$. Let us add a series of higher order interaction terms for the scalar to consider this matter further:
\begin{eqnarray}
S_{int}&=&\sum_{q=2}^\infty \int d^{7}x\sqrt{\bar{G}}\left\{\frac{\lambda_{2q}}{M_*^{4q-6}}\Phi^{2q}\right\}\delta(y-\pi r_c),\nonumber\\
&=&\sum_{q=2}^\infty\int d^{6}xe^{-6k\pi r_c}\left\{\frac{\lambda_{2q}}{M
_*^{4q-6}}e^{4q k\pi r_c}\Phi^{2q}\right\},\nonumber\\
&=&\sum_{q=2}^\infty\frac{\lambda_{2q}}{[e^{-k\pi r_c}M_*]^{4q-6}}\int d^{6}x\Phi^{2q},
\end{eqnarray}
where $\lambda_{2q}$ is a dimensionless coupling and we have performed the rescaling necessary to return the kinetic term to a canonical form to obtain the second line. The brane cutoff is warped down to $\Lambda_{IR}=e^{-k\pi r_c}M_*$, exactly as occurs in RS models. One may expect that for $R^{-1}>\Lambda_{IR}$ the IR brane theory could be trusted provided one neglects all KK modes whose mass exceeds the brane cutoff. However the effective description on the brane breaks down \emph{even when these states are discarded}, as is seen by considering the interactions involving only the zero modes. In the effective 4D theory these are
\begin{eqnarray}
S_{int}&=&\sum_{q=2}^\infty\frac{1}{\Lambda_{IR}^{2q-4}}\frac{\lambda_{2q}}{[e^{-k\pi r_c}M_*(2\pi R)]^{2q-2}}\int d^{4}x\phi^{(0)2q}+...\label{brane_int}
\end{eqnarray}
As an example consider $M_*\sim[2\pi R]^{-1}$, a relationship which, from the 7D perspective, appears within the range of validity of the effective theory description as $R^{-1}<M_*$.  However in the 4D theory the coupling of the zero mode quartic interaction $\phi^{(0)4}$ is $\sim\lambda_4 e^{2k\pi r_c}$ and if the 7D couplings assume `natural' values of order $\lambda_{2q}\in [10^{-2},1]$ this 4D coupling is severely non-perturbative. A similar enhancement is found for the higher order interaction terms. Thus it is not enough to simply discard the higher KK modes whose mass exceeds the IR brane cutoff; the effective theory description has broken down even for the zero modes. We deduce that the usual constraint  $R^{-1}<M_*$, required to ensure validity of the effective theory description, is inadequate to ensure that the effective 4D theory on the IR brane is sensible on a slice of $AdS_7$. This result holds more generally on a slice of $AdS_d$. 

The effective description of the 4D theory on the IR brane does make sense provided the transverse radius is less than the IR brane cutoff, $R^{-1}<\Lambda_{IR}= e^{-k \pi r_c}M_*$. This motivates the assumed value of $R^{-1}\lesssim e^{-k\pi r_c}M_*\sim$~TeV employed in the text. One may understand why the validity of the effective theory description requires  $R^{-1}<\Lambda_{IR}$ rather than simply $R^{-1}< M_*$ as follows. Consider two points with separation  $\Delta x^a \sim R$ in the transverse space. When localized on one of the branes this corresponds to a physical separation of:
\begin{eqnarray}
|\Delta s|=\left\{ \begin{array}{rll}\Delta x^a &\sim R&\mathrm{for\ UV\ localization,}\\ e^{-k\pi r_c}\Delta x^a &\sim e^{-k\pi r_c}R&\mathrm{for\ IR\ localization.}\end{array}\right. 
\end{eqnarray}
Within the present effective theory description one may only talk sensibly about proper distances satisfying $\Delta s^{-1}\lesssim k$, which translates into $R^{-1}\lesssim k$ for UV localization and $R^{-1}\lesssim e^{-k\pi r_c} k$ for IR localization. With $k\sim M_*$ the latter relation gives $R^{-1}\lesssim \Lambda_{IR}$ as promised. As the IR brane theory breaks down for $R^{-1}>\Lambda_{IR}$ the effective theory description employed in this work remains valid for the entire space only for $R^{-1}<\Lambda_{IR}$.

The relation  $R^{-1}\lesssim e^{-k\pi r_c}M_*$ may seem strange as in the limit $r_c\rightarrow \infty$ the effective description breaks down for any finite $R$. However this behaviour is understood as in the $r_c\rightarrow \infty$ limit the spacetime has a conical singularity, which is observed by noting that at the horizon the proper radius in the transverse directions shrinks to zero as $e^{-k\pi r_c}R$. The resolution of this singularity requires knowledge of the UV completion; for example the slice of $AdS_d$ may emerge from a more fundamental string theory. The presence of this singularity is known already in the literature and has been discussed in~\cite{Ponton:2000gi}, where a supergravity embedding of $AdS_6$ was considered to flush out possible ways to resolve it.

Before concluding we briefly consider the case of UV brane localization with $R^{-1}\sim$~TeV to show that the resulting effective 4D couplings can be highly suppressed. If the brane scalar $\Phi$ is instead localized on the UV brane of $AdS_7$ equation (\ref{brane_int}) becomes
\begin{eqnarray}
S_{int}^{\mathrm{UV}}&=&\sum_{q=2}^\infty\frac{1}{M_*^{2q-4}}\frac{\lambda_{2q}}{[M_*(2\pi R)]^{2q-2}}\int d^{4}x\phi^{(0)2q}+...,\label{brane_int_uv}
\end{eqnarray}
and with $e^{-k\pi r_c}M_* 2\pi R\sim 1$ and $\lambda_{2q}\sim1$ the effective zero mode coupling, for a given value of $q$, is:
\begin{eqnarray}
\sim e^{-2(q-1)k\pi r_c}M_*^{4-2q}.
\end{eqnarray}
As expected, the UV brane cutoff is $\Lambda_{UV}=M_*$ whilst the effective dimensionless coupling is of order $e^{-(2q-2)k\pi r_c}$, which, even for the zero mode quartic coupling case of $q=2$, is highly suppressed with $e^{-2k\pi r_c}\sim 10^{-15}$. A similar suppression holds more generally for $AdS_d$ and thus the SM cannot be localized on the UV brane for the phenomenologically interesting case of $R^{-1}\sim$~TeV. We also note that the effective coupling for the interaction $\phi^{(0)2}\phi^{(n_a)2}$ between two zero modes and two $n_a\ne0$ KK modes is $\sim e^{-2k\pi r_c}$ ($\sim 1$) in the case of UV (IR) localization when $\lambda_4\sim\mathcal{O}(1)$. It is easy to understand why the effective quartic couplings on the IR brane can be $\mathcal{O}(1)$ whilst those on the UV brane must be highly suppressed. The running of the quartic coupling will receive contributions from loops containing transverse KK mode scalars and, if the relevant effective 4D couplings are of order unity, will rapidly become non-perturbative. This is ok on the IR brane where the cutoff is warped down to $e^{-k\pi r_c}M_*\sim$~TeV so that a rapid approach to the non-perturbative regime is consistent with the expectation that the IR brane theory will break down at the TeV scale. However on the UV brane the description is expected to be valid up to the fundamental scale $M_*$. This requires the effective 4D couplings to be highly suppressed to ensure a slow running and to avoid a breakdown of the theory at scales $E\ll M_*$. In this way we observe that the theory automatically generates couplings that are appropriate for, and consistent with, the expected domain of validity of the effective theory description when $R^{-1}\lesssim e^{-k\pi r_c}M_*$. 
\section{Conclusion\label{sec:conc}}
In this work we have extended the program begun in~\cite{McDonald:2009md} for $AdS_5\times T^2$ and considered the promotion of the RS model to a higher dimensional slice of $AdS_d$ for $d>5$. Such spaces are interesting as they admit a generalized version of the warped realization of the weak scale employed in the RS model. Our primary motivation was to determine the viability of combining the phenomenologically appealing features of RS and UED models in such spaces. We have performed the KK reduction for gravitons, bulk vectors and, for the case of $AdS_6$, the gauge-scalars. We also obtained the KK spectra for bulk fermions on a slice of $AdS_{7,9}$ and showed that the RS approach to flavor generalizes to these spaces with the localization of chiral zero mode fermions controlled by their bulk Dirac mass parameters. However for the phenomenologically interesting case where the transverse radius is $R^{-1}\sim$~TeV we find that bulk standard model fields are not viable due to a resulting volume suppression of the gauge coupling constants. A similar suppression occurs for UV localization so that, when propagating in the transverse directions, the SM fields should be confined to the IR brane, consistent with the warped realization of the weak/Planck hierarchy. The main experimental signature of the $AdS_d$ spaces in this instance is the observation of warped KK gravitons in addition to the usual UED KK modes.
\section*{Acknowledgments}
The author thanks D. Morrissey and gives special thanks to B. Batell (for many communications during the formative stages of this work)  and H. Davoudiasl (for comments on the discussion of Section~\ref{sec:eff_4d_theory}). The author also thanks the Perimeter Institute for kind hospitality whilst part of this work was undertaken. This work was supported by the Natural Science and Engineering Research Council of Canada and in part by the Perimeter Institute for Theoretical Physics. Research at Perimeter Institute is supported by the Government of Canada through Industry Canada and by the Province of Ontario through the Ministry of Research and Innovation.
\appendix
\section*{Appendix}
\section{Embedding $\mathbf{AdS_{d-1}}$ in $\mathbf{AdS_d}$\label{app:embedding}}
The Poincare parametrization of the $AdS_d$ metric is:
\begin{eqnarray}
ds^2_{AdS_d}&=&\frac{1}{(kz)^2}\left[\eta_{\mu\nu}dx^\mu dx^\nu -\delta_{ab}dx^{a}dx^{b}-dz^{2}\right],
\end{eqnarray}
and although the geometry is non-factorizable it may be expressed in terms of an embedded $AdS_{d-1}$ by changing coordinates to
\begin{eqnarray}
x^{d-1}=Z \cos\phi\quad,\quad z=Z\sin\phi,
\end{eqnarray}
to obtain
\begin{eqnarray}
ds^2_{AdS_d}&=&\frac{1}{\sin^2\phi^2}\left[ds^2_{AdS_{d-1}}-k^{-2}d\phi^{2}\right],
\end{eqnarray}
where the $AdS_{d-1}$ metric is:
\begin{eqnarray}
ds^2_{AdS_{d-1}}&=&\frac{1}{(kZ)^2}\left\{\eta_{\mu\nu}dx^\mu dx^\nu -\sum_{a=5}^{d-2}(dx^{a})^2-dZ^2\right\}.
\end{eqnarray}
Thus the warped direction for the embedded $AdS_{d-1}$ differs from that of the original $AdS_d$~\cite{Metsaev:2000qb}. One can repeat this process to obtain an embedding of $AdS_5$ in $AdS_d$. Consequently the warped profiles in the KK decomposition of bulk fields on a compactified slice of $AdS_d$ are not expected to reduce to the familiar $AdS_5$ expressions. 
\section{Fermions in 7D\label{app:ferm_not}}
The generators of the 7D Lorentz group $SO(1,6)$ for the spin 1/2 representation are
\begin{eqnarray}
S^{MN}=\frac{\Sigma^{MN}}{2}=\frac{i}{4}[\Gamma^M,\Gamma^N],
\end{eqnarray}
where the 7D gamma matrices satisfy
\begin{eqnarray}
\{\Gamma^M,\Gamma^N\}=2\eta^{MN}I,\label{clifford_algebra}
\end{eqnarray}
and $\eta^{MN}=\mathrm{diag}(1,-1,-1,...)$. Note that in 7D the minimum dimensionality of the matrices satisfying the Clifford algebra (\ref{clifford_algebra}) is $8\times8$ so that fermions are described by spinors with eight components. We employ the 7D generalization of the of the Weyl representation for the $\Gamma$-matrices. For $M=0,1,2,3,5,6$ we have
\begin{eqnarray}
\Gamma^M=\left(\begin{array}{cc}0&\Sigma^M\\\bar{\Sigma}^M&0\end{array}\right)\label{6D_gamma},
\end{eqnarray}
where 
\begin{eqnarray}
\Sigma^0&=&\bar{\Sigma}^0=\gamma^0\gamma^0\quad,\quad\Sigma^i=-\bar{\Sigma}^i=\gamma^0\gamma^i\\
\Sigma^5&=&-\bar{\Sigma}^5=i\gamma^0\gamma^5\quad,\quad \Sigma^6=-\bar{\Sigma}^6=\gamma^0,
\end{eqnarray}
and for definiteness we employ the Weyl representation of the Dirac gamma matrices
\begin{eqnarray}
\gamma^0&=&\left(\begin{array}{cc}0&1\\1&0\end{array}\right)\quad,\quad\gamma^i=\left(\begin{array}{cc}0&\sigma^i\\-\sigma^i&0\end{array}\right)\quad,\quad\gamma^5=\left(\begin{array}{cc}-1&0\\0&1\end{array}\right).\label{dirac_matrices_weyl}
\end{eqnarray}
In 4D the projection operators $P_{R,L}=\frac{1}{2}(1\pm\gamma^5)$ project out the right- and left-chiral components of a Dirac spinor. These operators may be generalized to 7D as
\begin{eqnarray}
 P^7_{R,L}&=&\frac{1}{2}(1\pm i\Gamma^0\Gamma^1\Gamma^2\Gamma^3).
\end{eqnarray}
The final gamma matrix is
\begin{eqnarray}
\Gamma^7&=&i\bar{\Gamma}\equiv i\Gamma^0\Gamma^1\Gamma^2\Gamma^3\Gamma^5\Gamma^6=i\left(\begin{array}{cc}-I&0\\0&I\end{array}\right),
\end{eqnarray}
which may be used to define the projection operators
\begin{eqnarray}
 P_\pm&=&\frac{1}{2}(1\pm\bar{\Gamma}).
\end{eqnarray}
Thus one may label the components of the 7D spinor with their 6D chirality ($\pm$) and their 4D chirality ($R,L$) as
\begin{eqnarray}
\Psi =\left(\psi_{-R},\psi_{-L},\psi_{+L},\psi_{+R}\right)^T.\label{7d_fermi}
\end{eqnarray}
\section{Fermions in 9D\label{app:ferm_not_9}}
The generators of the 9D Lorentz group $SO(1,8)$ for the spin 1/2 representation are
\begin{eqnarray}
S^{MN}=\frac{\Sigma^{MN}}{2}=\frac{i}{4}[\mathcal{G}^M,\g^N],
\end{eqnarray}
with
\begin{eqnarray}
\{\g^M,\g^N\}=2\eta^{MN}I.\label{clifford_algebra_9}
\end{eqnarray}
In 9D fermions are described by spinors with sixteen components. We employ a generalized Weyl representation of the $\g$-matrices, which, for $M\ne9$, may be written in terms of the 7D Dirac matrices as
\begin{eqnarray}
\g^M=\left(\begin{array}{cc}0&\Omega^M\\\bar{\Omega}^M&0\end{array}\right)\label{9D_gamma},
\end{eqnarray}
where 
\begin{eqnarray}
\Omega^0&=&\bar{\Omega}^0=\Gamma^0\Gamma^0\quad,\quad\Omega^i=-\bar{\Omega}^i=\Gamma^0\Gamma^i\quad,\quad\Omega^5=-\bar{\Omega}^5=\Gamma^0\Gamma^5\\
\Omega^6&=&-\bar{\Omega}^6=\Gamma^0\Gamma^6\quad,\quad \Omega^7=-\bar{\Omega}^7=\Gamma^0\Gamma^7\quad,\quad \Omega^8=-\bar{\Omega}^8=\Gamma^0.
\end{eqnarray} 
In 4D the projection operators $P_{R,L}=\frac{1}{2}(1\pm\gamma^5)$ project out the right- and left-chiral components of a Dirac spinor. These operators may be generalized to 9D as
\begin{eqnarray}
 P^9_{R,L}&=&\frac{1}{2}(1\pm i\g^0\g^1\g^2\g^3),
\end{eqnarray}
and the 6D projection operators $P_\pm=\frac{1}{2}(1\pm\bar{\Gamma})$ also generalize to the 9D operators
\begin{eqnarray}
 P^9_{\pm}&=&\frac{1}{2}(1\pm\g^0\g^1\g^2\g^3\g^5\g^6).
\end{eqnarray}
The final gamma matrix is
\begin{eqnarray}
\g^9&=&i\bar{\g}\equiv \g^0\g^1\g^2\g^3\g^5\g^6\g^7\g^8=i\left(\begin{array}{cc}-I_{8\times8}&0\\0&I_{8\times8}\end{array}\right),
\end{eqnarray}
which may be used to define the projection operators
\begin{eqnarray}
 P_{\uparrow,\downarrow}&=&\frac{1}{2}(1\pm\bar{\g}).
\end{eqnarray}
Thus one may label the components of the 9D spinor with their 8D chirality ($\uparrow,\downarrow$) as 
\begin{eqnarray}
\Psi =\left(\begin{array}{c} \psi_\downarrow\\\psi_\uparrow\end{array}\right),\label{9d_16_fermi}
\end{eqnarray}
and one can further label the components of $\psi_{\uparrow,\downarrow}$ by their 6D chirality ($\pm$) and their 4D chirality ($R,L$) as
\begin{eqnarray}
\psi_\downarrow=\left(\psi_{\downarrow+L},\psi_{\downarrow+R},\psi_{\downarrow-R},\psi_{\downarrow-L}\right)^T\quad,\quad\psi_\uparrow=\left(\psi_{\uparrow-R},\psi_{\uparrow-L},\psi_{\uparrow+L},\psi_{\uparrow+R}\right)^T.\label{9d_fermi}
\end{eqnarray}
The above notation clearly labels the components of $\Psi$ in terms of their various lower dimensional chiral properties. It is, however, somewhat cumbersome and we employ a simpler notation in the text; see (\ref{9d_fermi_compon_label}).
\section{Fermion Wave functions \label{app:t_wavef}}
\subsection{Toroidal wave functions: $\mathbf{AdS_7}$\label{app:t_wave_7}}
The fermion wave functions on the toroidal dimensions may be written in terms of $g^{(n_a)}_{+(-)}$, the usual expansions for the even (odd) KK modes on the $T^2/Z_2'$ orbifold:
\begin{eqnarray}
g^{(n_a)}_+ (x^a)&=&\frac{1}{\sqrt{2}\pi R} \left(\frac{1}{\sqrt{2}}\right)^{\delta_{n_a 0}}\cos\left[\frac{n_5 x^5+n_6x^6}{R}\right],\label{ued_+_profile}\\
g^{(n_a)}_- (x^a)&=&\frac{1}{\sqrt{2}\pi R} \sin\left[\frac{n_5 x^5+n_6x^6}{R}\right],\label{ued_-_profile}
\end{eqnarray}
where $n_a=(n_5,n_6)$. For \ads with toroidal compactification the $T^2$ profiles must satisfy
\begin{eqnarray}
(\partial_5\pm i\partial_6)g^{(n_a)}_{+ L,R}&=& \mp m_{n_a}g^{(n_a)}_{+R,L},\\
(\partial_5\mp i\partial_6)g^{(n_a)}_{- L,R}&=&\pm m_{n_a}g^{(n_a)}_{-R,L},
\end{eqnarray}
and may be written as
\begin{eqnarray}
g^{(n_a)}_{+ L}(x^a)&=&g^{(n_a)}_{-R}(x^a)=g^{(n_a)}_+ (x^a),\\
g^{(n_a)}_{+ R}(x^a)&=&g^{(n_a)}_{- L}(x^a)= \frac{n_5+in_6}{\sqrt{n_5^2+n_6^2}}g^{(n_a)}_- (x^a).
\end{eqnarray}
\subsection{Toroidal wave functions: $\mathbf{AdS_9}$\label{t_wave_9}}
The wave functions along $x^a$ must satisfy
\begin{eqnarray}
(\partial_5\pm i\partial_6)g^{(n_a)}_{\alpha L,R}&=& \mp m_{n_a}g^{(n_a)}_{\alpha R,L}\quad\mathrm{for}\quad\alpha=1,4,\\
(\partial_5\mp i\partial_6)g^{(n_a)}_{\alpha L,R}&=&\pm m_{n_a}g^{(n_a)}_{\alpha R,L}\quad\mathrm{for}\quad\alpha=2,3,
\end{eqnarray}
giving
\begin{eqnarray}
g^{(n_a)}_{1 L}(x^a)&=&g^{(n_a)}_{2R}(x^a)=g^{(n_a)}_{3R}(x^a)=g^{(n_a)}_{4L}(x^a)=g^{(n_a)}_+ (x^a),\\
g^{(n_a)}_{1 R}(x^a)&=&g^{(n_a)}_{2 L}(x^a)=g^{(n_a)}_{3 L}(x^a)=g^{(n_a)}_{4 R}(x^a)= \frac{n_5+in_6}{\sqrt{n_5^2+n_6^2}}g^{(n_a)}_- (x^a),
\end{eqnarray}
where we express the solutions in terms of (\ref{ued_+_profile}), (\ref{ued_-_profile}). Similarly the wave functions along $x^b$ satisfy
\begin{eqnarray}
(\partial_7\pm i\partial_8)h^{(n_b)}_{\alpha R,L}&=& \pm m_{n_b}h^{(n_b)}_{\beta L,R}\quad\mathrm{for}\quad(\alpha,\beta)=(1,2),(3,4),\\
(\partial_7\mp i\partial_8)h^{(n_b)}_{\alpha R,L}&=& \mp m_{n_b}h^{(n_b)}_{\beta L,R}\quad\mathrm{for}\quad(\alpha,\beta)=(2,1),(4,3),
\end{eqnarray}
where $n_b=(n_7,n_8)$. The solutions are
\begin{eqnarray}
h^{(n_b)}_{2 L}(x^b)&=&h^{(n_b)}_{2 R}(x^b)=h^{(n_b)}_{4L}(x^b)=h^{(n_b)}_{4R}(x^b)=g^{(n_b)}_+ (x^b),\\
h^{(n_b)}_{1 L}(x^b)&=&h^{(n_b)}_{1 R}(x^b)=h^{(n_b)}_{3L}(x^b)=h^{(n_b)}_{3R}(x^b)= \frac{n_7-in_8}{\sqrt{n_7^2+n_8^2}}g^{(n_b)}_- (x^b),
\end{eqnarray}
with $g^{(n_b)}_\pm (x^b)$ given by (\ref{ued_+_profile}), (\ref{ued_-_profile}) with the replacement $n_a,x^a\rightarrow n_b,x^b$.
\subsection{Normalization factors in $\mathbf{AdS_9}$\label{app:norm}}
The equations of motion require the normalization factors for $f_{\alpha L,R}^{(\vec{n})}$ to be related and one can show that they may be expressed in terms of a single normalization factor $N_{\Psi}^{(\vec{n})}$ via
\begin{eqnarray}
N_{1L,R}^{(\vec{n})} &=& \frac{2\sqrt{m_{\vec{n}}(m_{\vec{n}}+ m_{n_b})}}{m_{\vec{n}}+ m_{n_a}+m_{n_b}}N_{\Psi}^{(\vec{n})},\\
N_{2L,R}^{(\vec{n})} &=& \frac{2\sqrt{m_{\vec{n}}(m_{\vec{n}}+ m_{n_b})}}{m_{\vec{n}}- m_{n_a}+m_{n_b}}N_{\Psi}^{(\vec{n})},\\
N_{3L,R}^{(\vec{n})} &=& N_{4L,R}^{(\vec{n})} =2\sqrt{\frac{m_{\vec{n}}(m_{\vec{n}}+ m_{n_b})}{m_{\vec{n}}^2- m_{n_a}^2-m^2_{n_b}}}N_{\Psi}^{(\vec{n})}.
\end{eqnarray}

\end{document}